\documentclass{article}
\usepackage{listings}
\usepackage{graphicx}
\usepackage{amssymb}
\usepackage{pstricks}
\usepackage{color}
\usepackage{hyperref} 
\usepackage{verbatim}
\usepackage{makeidx}
\usepackage{amsfonts}
\usepackage{amsmath}
\makeindex
\large \normalsize
\newtheorem{theorem}{Theorem}[section]

\newtheorem{lemma}[theorem]{Lemma}
\newtheorem{proposition}{Proposition}

\def \epsilon {\varepsilon}
\def\norm#1{\left |#1\right |}
\def\Norm#1{\left \|#1\right \|}

\title{Evolution of the tangent vectors and localization\\ 
of the stable and unstable manifolds of hyperbolic orbits\\ 
by Fast Lyapunov Indicators}

\author{Massimiliano Guzzo\\ 
Dipartimento di Matematica\\ Via Trieste, 63 - 35121 Padova, Italy \\ 
guzzo@math.unipd.it\\
\\
Elena Lega\\
Universit\'e de Nice Sophia Antipolis, CNRS UMR 7293\\
Observatoire de la C\^ote d'Azur\\  
Bv. de l'Observatoire, B.P.~4229,  06304 Nice cedex 4, France\\
elena.lega@oca.eu
}

\date{  }
\begin{document}
\maketitle

\begin{abstract} \par \noindent
The Fast Lyapunov Indicators are functions defined on the tangent fiber of the 
phase--space of a discrete (or continuous) dynamical system, by using a finite 
number of iterations of the dynamics. In the last decade, they have been 
largely used in numerical computations to localize the resonances in 
the phase--space and, more recently, also the stable and unstable manifolds of 
normally hyperbolic invariant manifolds. In this paper, we provide an  
analytic description of the growth of tangent vectors for orbits with 
initial conditions which are close to the stable-unstable manifolds of a 
hyperbolic saddle point of an area--preserving map. The representation explains why 
the Fast Lyapunov Indicator detects the stable-unstable manifolds of all fixed 
points which satisfy a certain condition. If the condition is not satisfied, 
a suitably modified Fast Lyapunov Indicator can be still used to detect the 
stable-unstable manifolds. The new method allows for a detection of the 
manifolds with a number of precision digits which increases linearly with 
respect to the integration time. We illustrate the method on the critical 
problem of detection of the so--called tube manifolds of the Lyapunov orbits of 
$L_1,L_2$ in the circular restricted three--body problem. 
\end{abstract}

\section{Introduction}

Since the first detection of chaotic motions in 1964 (Henon--Heiles 
\cite{henheil}), several indicators have been largely used to characterize the 
different dynamics of dynamical systems. Many dynamical indicators,  such as 
the Lyapunov characteristic exponents and the more 
recently introduced finite--time chaos indicators (such as the 
Finite Time Lyapunov Exponent--FTLE \cite{tb}, Fast 
Lyapunov Indicator--FLI \cite{FroLeGo96}, Mean Exponential Growth of 
Nearby Orbits--MEGNO \cite{cs}),
are defined by the local divergence of nearby initial conditions, that is
by the variational dynamics. For example, for a discrete dynamical system defined by the map
\begin{eqnarray}
\Phi: M &\longrightarrow&  M\cr
z &\longmapsto& \Phi(z)   ,
\end{eqnarray}
with $M\subseteq {\Bbb R}^{n}$ open invariant, by denoting with $D\Phi_z$ the 
tangent map of $\Phi$ at $z$:
\begin{eqnarray}
D\Phi_z: {\Bbb R}^{n} &\longrightarrow&  {\Bbb R}^{n}\cr
v &\longmapsto&   D\Phi_z v ,
\end{eqnarray}
the characteristic Lyapunov exponent of a point $z\in M$ and a vector 
$v\in  {\Bbb R}^n\backslash 0$ is defined by the limit
\begin{equation}
\lambda(z,v)= \lim_{T\rightarrow +\infty} {1\over T} 
\log {\Norm{D\Phi^T_z v}\over \Norm{v}}   ,
\label{lyapexp}
\end{equation}
and the largest Lyapunov exponent of $z$ is the maximum
of $\lambda(z,v)$ for $v \ne 0$. As a matter of fact, the numerical 
estimation of the characteristic Lyapunov exponents (see \cite{BenGalStre76}) 
relies on extrapolation of finite time computations, since computers cannot 
integrate on infinite time intervals.  The so--called finite--time chaos 
indicators (such as the FTLE, the FLI and the MEGNO) have been afterwards 
introduced as surrogate indicators of the largest Lyapunov exponent, 
 with the aim to discriminate between regular orbits and chaotic orbits 
using time intervals which are significantly smaller than the time 
interval required for a reliable estimation of the largest characteristic
 Lyapunov exponent (\cite{FroLeGo96}, \cite{cs}). For 
example, the function Fast Lyapunov Indicator of $z$ and $v$ is simply 
defined by
\begin{equation}
l_T(z,v )= \log {\Norm{D\Phi^T_z v}\over \Norm{v}} ,
\label{fli}
\end{equation}
and depends parametrically on the integer $T>0$, as well as on the choice 
of a norm on ${\Bbb R}^{n}$. The definition of finite time chaos indicators was
justified by the possibility of their systematic numerical computation over 
large grids of initial conditions in the phase--space in a reasonable 
computational time.  We remark that, specifically in Celestial Mechanics, the 
numerical detection of the resonances of a system using dynamical indicators, 
both formulated using the Lyapunov exponent theory or alternatively the 
Fourier analysis (such as the frequency 
analysis \cite{Laskar90,Laskar93,Laskar92}), 
is one of the major tools for studying its long--term instability (for recent 
examples, see \cite{rl,rg,robutel,mf,guzzoplanets,guzzoplanets2,wmd}).
The papers \cite{FGL00},\cite{GLF02}, focused and proved 
properties of the finite time chaos indicators, specifically the FLI, 
which are lost by taking the limit of $l_T(z,v )/T$, thus differentiating 
the use of these indicators from the parent largest Lyapunov characteristic exponent. Specifically, since \cite{FGL00},\cite{GLF02}, the FLI has been 
used to discriminate regular motions of different nature: for example 
the motions which are regular because are supported by a KAM torus from 
the regular motions in the resonances of a system. This property of the FLI 
improved a lot the precision in the numerical localization of different 
types of resonant motions, the so--called Arnold web, and provided the 
technical tool for the first numerical computations 
of diffusion along the resonances of quasi--integrable systems in 
exponentially long times \cite{LGF03,GLF05,FGL05,GLF11, GL13}, as depicted 
in the celebrate Arnold's paper \cite{Arnold64}.  

More recently, the FLI has 
been successfully used to compute the stable and unstable manifolds of 
normally hyperbolic invariant manifolds of the standard map
and its generalizations \cite{Guzzob07,GLF09Fiori}, and of the   
three--body--problem \cite{Villac08,LGF10b,GL13mnras}. In these cases  
it happens that, depending on the choice of the parameter $T$, finite 
pieces of the stable and unstable manifolds appear as sharp local maxima of the 
FLI. As a matter of fact, the possibility of sharp detection of the stable 
and unstable manifolds of a fixed point, or periodic orbit, with a FLI 
computation is not general and turns out to be a property of the manifolds. 
A model example is represented by the stable and unstable manifold 
of the fixed point $(0,0)$ of the symplectic map
\begin{equation}
\Phi(\varphi ,I)=\left (\varphi +I\ ,\ I+ {\sin (\varphi +I)\over 
(\sigma \cos(\varphi +I)+2)^2} \right ) ,
\label{modelmap}
\end{equation}
where $(\varphi,I)\in M=(2\pi {\Bbb S}^1)\times {\Bbb R}$ are the phase--space
variables, $\sigma =\pm 1$ is a parameter: for $\sigma=-1$ the 
FLI may be used for excellent detection 
of the manifolds; for $\sigma=1$ the FLI does not provide any detection.  

To explain this fact, in this paper we provide a representation for 
the growth of tangent vectors for orbits with initial conditions close to the 
stable manifold of a saddle fixed point. To better illustrate the theory, 
we consider a two dimensional area--preserving map with a saddle fixed point
$z_*$, but the techniques which we use (the local stable manifold theorem 
and Lipschitz estimates) can be used also in 
the higher dimensional cases. The two dimensional case allows us to treat 
also Poincar\'e sections of the circular restricted three body problem. 

Let us denote by $z_*$ the saddle point of the map, and by 
$W_s,W_u$ its stable and unstable manifold. We consider a point $z_s\in W_s$, 
a tangent vector $v\in {\Bbb R}^2$, and we provide estimates about the norm 
of the tangent vector $D\Phi^T_z v$, for points $z\notin W_s$ which are 
close to $z_s$. As it is usual, the same arguments applied to the inverse 
map $\Phi^{-1}$, allow to reformulate the result by exchanging the role of the 
stable manifold with that of the unstable manifold. For the points $z$ which are the suitably close to $z_s\in W_s$, the orbit  $\Phi^k(z)$ follows 
closely the orbit $\Phi^k(z_s)$ for any $k\leq T$, and $\Norm{D\Phi^k_z v}$ 
remains close to $\Norm{D\Phi^k_{z_s} v}$ as well. The most interesting 
situation happens for the points $z$ which are little more distant from the 
stable manifold: their orbit (i) follows closely the orbit $\Phi^k(z_s)$ 
only for $k$ smaller than some $K_0 < T$; (ii) then remains close to the 
hyperbolic fixed point (for a number of iterations which increases 
logarithmically with respect to some distance between $z$ and $z_s$, see Section
\ref{Section2}), (iii) then follows closely the orbit of a point on the 
unstable manifold $W_u$ in the remaining iterations. It is during 
the process (iii) that the growth of the tangent vector 
$\Norm{D\Phi^k_z v}$ can be significantly different from the growth of 
$\Norm{D\Phi^k_{z_s} v}$, and the difference may be possibly used to 
characterize the distance of $z$ from  the stable manifold. As a matter of fact, 
with evidence any difference may exist only due to the non--linearity of the 
map $\Phi$.  In Section \ref{Section2} we provide a representation 
for such a difference, and we discuss a condition which guarantees the 
desired scaling of the FLI with 
respect to the distance of $z$ from the stable manifold. If this condition is
satisfied, the computation of the FLI on a grid of initial conditions provides 
a sharp detection of the stable and unstable manifolds (see Section 
\ref{Section3}): typically, the time $T$ used for the FLI computation,  
which is the time needed by the orbits with initial condition $z$ to approach 
the fixed point $z_*$, turns out to be proportional 
to the number of precision digits of the detection. 

At the light of the representation provided in Section \ref{Section2}, 
we propose a generalization of the FLI which weakens a lot the 
condition for the detection of the stable and unstable manifold. For any
smooth and positive function 
$$
u : {\Bbb R}^2\rightarrow {\Bbb R}^+
$$
we define the modified FLI indicator of 
$z\in M$, $v\in {\Bbb R}^2$ at time $T>0$, as the $T$--th element of the 
sequence
\begin{equation}
l_{1}=\ln \Norm{v}\ \ ,\ \ 
l_{j+1}=l_j + u({ z_j}) \ln { \Norm{D\Phi_{z_j} v_j}\over \Norm{v_j}} ,
\label{ltind}
\end{equation}
{ where $z_j:=\Phi^j(z)$ and $v_j:=D\Phi^j_{z_j}v$}.
The traditional FLI is obtained with the choice $u(z)=1$ for any $z\in M$. 
We consider the alternative case of functions $u(z)$ which are test functions 
of some neighbourhood ${\cal B}\subseteq M$ of the fixed point, 
 and precisely with
$u(z)=1$ for $z\in {\overline {\cal B}}$, and $u(z)=0$ for $z$ outside a given 
open set $V \supseteq {\overline {\cal B}}$. When the diameter of the 
set ${\cal B}$ is small, but not necessarily extremely small, the 
computation of the modified FLI indicator 
allows to refine the localization of the fixed point by many orders
of magnitude. Therefore, 
at variance with the traditional FLI indicator, the modified indicators 
are proposed as a general tool for the numerical detection of the stable and 
unstable manifolds. An illustration of the potentialities of these indicators
is given in Section 3, where we  provide computations of the stable and 
unstable manifolds and their heteroclinic intersections, of the Lyapunov 
orbits around $L_1$, $L_2$ of the circular restricted three--body problem. The 
application is particularly critical, since these  
manifolds are located in a region of the phase--space close to the 
singularity due to the secondary mass.  

The paper is structured as follows. In Section 2 we provide the  
representation for the evolution $D\Phi^T_z v$ of the norm of tangent 
vector $v$ for points $z\notin W_s$ which are 
suitably close to the stable manifold, and we also discuss  a sufficient 
condition for the FLI to detect sharply the stable and unstable manifolds of 
the map. In Section 3 we provide an illustration of the method for the 
computation of the stable and 
unstable manifolds of the Lyapunov 
orbits around $L_1$, $L_2$ of the circular restricted three body problem; 
in Section 4 we provide the proof of Proposition \ref{pro1}. In Section 5 we 
formulate and prove two technical lemmas. Finally, Conclusions are provided 
in Section 6.

\section{Evolution of the tangent vectors close to the stable manifolds of the 
saddle points of two dimensional area--preserving maps}\label{Section2}

We consider a smooth two--dimensional area--preserving map:
\begin{equation}
\Phi(z) = A z +f(z) ,
\label{eqipbi}
\end{equation}
where $A$ is a $2\times 2$ diagonal matrix with $A_{11}=\lambda_u>1$,
$A_{22}=1/\lambda_u$ and  $f$ is at least quadratic in $z_1,z_2$, 
that is $f_i(0,0)=0$ and ${\partial f_{i}\over\partial z_j}(0,0)=0$, for 
any $i,j=1,2$. Therefore, the origin is a saddle fixed point. 

We need to introduce some constants which characterize the analytic properties
of $\Phi$. We denote by $\lambda_\Phi,\lambda_{\Phi^{-1}},\lambda_{D\Phi}$ 
the Lipschitz constants of $\Phi,\Phi^{-1},D\Phi$ respectively defined with 
respect to the norm $\Norm{u} :=\max\{ \norm{u_1},\norm{u_2}\}$,
in the set $B(R)=\{ z:\  \Norm{z}\leq R\}$. Also, we set 
$\eta$ such that, for any $z\in B(R)$, we have
$$
\Norm{f(z)}\leq \eta\Norm{z}^2\ \ ,\ \ 
\Norm{Df_z} \leq \eta \Norm{z}\ \ ,\ \ 
\Norm{D^2f_z}\leq \eta
$$
$$
\Norm{f(z')-f(z'')}\leq \eta \max \{ \Norm{z'},\Norm{z''} \}\Norm{z'-z''}  ,
$$
where $D^2f_z$ denotes the Hessian matrix of $f$ at the point $z$ and, 
by denoting with $\Phi^{-1}(z)=A^{-1}z+ {\tilde f}(z)$ the inverse map, 
we also have
$$
\Norm{\tilde f(z)}\leq \eta\Norm{z}^2\ \ ,\ \ 
\Norm{D{\tilde f}_z} \leq \eta \Norm{z}\ \ ,\ \ 
\Norm{D^2{\tilde f}_z}\leq \eta
$$
$$
\Norm{{\tilde f}(z')-{\tilde f}(z'')}\leq \eta \max 
\{ \Norm{z'},\Norm{z''} \}\Norm{z'-z''}  .
$$
Moreover, since $\Phi$ is a diffeomorphism, we have
\begin{equation}
\sigma= \min_{z\in B(R)}\min_{\Norm{v}=1}\Norm{D\Phi_z v} > 0  .
\label{sigma}
\end{equation}
By the local stable manifold theorem, we consider e neighbourhood 
$B(r_*)$ of the origin where the local stable and 
unstable manifolds $W^l_s,W^l_u$ are Cartesian graphs over the $z_2$ and 
$z_1$ axes respectively, that is
$$
W^l_s =\left \{ z:  \norm{z_2}\leq r_*\ \ ,\ \ z_1=w_s(z_2)\right \}
$$
$$
W^l_u =\left \{ z:  \norm{z_1}\leq r_*\ \ ,\ \ z_2=w_u(z_1)\right \}
$$
with $w_s(0)=w_u(0)=0$, $w'_s(0)=w'_u(0)=0$ and, by possibly increasing $\eta$, 
$$
\norm{w_s(z_2)}\leq \eta \norm{z_2}^2\ \ ,\ \ 
\norm{w_u(z_1)}\leq \eta \norm{z_1}^2  
$$
and
$$
\norm{w_s(\xi')-w_s(\xi'')}\leq  \lambda_w \max\{ \norm{\xi'},\norm{\xi''} \}\norm{\xi'-\xi''}
$$
$$
\norm{w_u(\xi')-w_u(\xi'')}\leq  \lambda_w \max\{ \norm{\xi'},\norm{\xi''}\}\norm{\xi'-\xi''}  .
$$
We denote by $W_s,W_u$ the stable and unstable manifolds of the 
origin. We consider a point $z_s\in W_s$, a tangent vector 
$v\in {\Bbb R}^2$, and we provide estimates about the norm of the tangent 
vector $D\Phi^T_z v$, for points $z\notin W_s$ which are suitably 
close to $z_s$, precisely in a curve $z_\epsilon$, with $z_0=z_s$ and
$\Norm{z-z_\epsilon}=\epsilon$.

\begin{figure*}[!]
\begin{center}
\includegraphics[height=6cm,width=8cm]{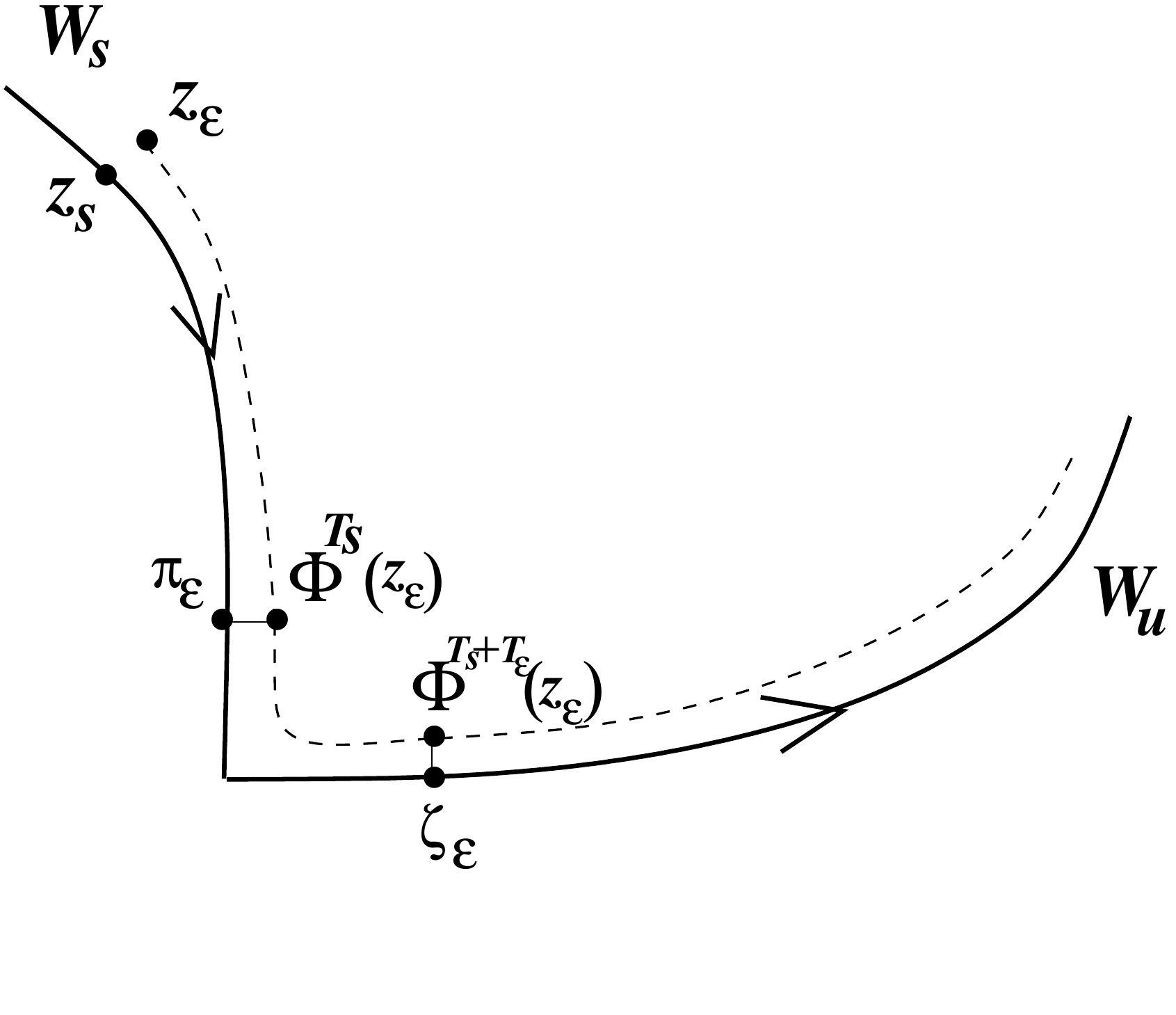}
\end{center}
\caption{Illustration of $z_s,z_\epsilon$; of $\Phi^{T_s}(z_\epsilon)$ and 
its parallel projection $\pi_\epsilon$ on the local stable manifold; of 
$\Phi^{T_s+T_\epsilon}(z_\epsilon)$ and its parallel projection $\zeta_\epsilon$ on 
the local unstable manifold. 
}
\label{deltas}
\end{figure*}

Let us consider a small $\delta := {\delta_0 \over T}$, with 
$\delta_0$ satisfying
$$
\delta_0 \leq \min \left ({1\over 16 \max (1,\eta )^2 e^3\lambda_u^2}\left ( 
1-{1\over \lambda_u}\right ) , {r_*\over 2}\right ) .
$$
Then, we consider the minimum $T_s:=T_s(\delta)$ such that 
$\Phi^{T_s}(z_s)\in B(\delta-2\delta^2)$. Typically, one has 
$T_s\sim \ln(1/\delta)$. For all $\epsilon$, we have (see Lemma 
\ref{lemma2}):
\begin{equation}
\Norm{\Phi^{T_s}(z_\epsilon)-\Phi^{T_s}(z_s)}\leq \lambda_{\Phi}^{T_s}\epsilon
\end{equation}
\begin{equation}
\Norm{D\Phi^{T_s}_{z_\epsilon}v-D\Phi^{T_s}_{z_s}v}\leq \Norm{D\Phi^{T_s}_{z_s}v}
\lambda^{T_s}\epsilon  ,
\label{vtanA1}
\end{equation}
where $\lambda=\max (\lambda_\Phi, (\Norm{D\Phi}+\lambda_{D\Phi})/\sigma)$. 
We consider only the small $\epsilon$ satisfying 
$\lambda^{T_s}\epsilon < \delta^2$, so that $\Phi^{T_s}(z_\epsilon)\in B(\delta-\delta^2)$, are close to $\Phi^{T_s}(z_s)$ and $\Norm{D\Phi^{T_s}_{z_\epsilon}v}$ are close
to $\Norm{D\Phi^{T_s}_{z_s}v}$. We rename the vector $D\Phi^{T_s}_{z_s}v$
as follows:
$$
w=w_s+w_u=D\Phi^{T_s}_{z_s}v  ,
$$ 
where $w_s,w_u$ are the orthogonal projections of $w$ over the stable and 
unstable spaces of the matrix $A$,  i.e. the $z_2$ and $z_1$ axes, 
respectively. We need a condition which ensures that $v$ is not close to 
some special contracting direction. Precisely, we assume that the initial 
vector $v$ is such that
$$
\Norm{w_s}\leq \Norm{w_u} = \Norm{w}  .
$$
In particular, for any $k\geq 0$, we have $\Norm{A^kw} = \lambda_u^k \Norm{w_u}$.
\vskip 0.2 cm
\noindent
Let us denote by 
$$
\pi_\epsilon = \Big(w_s(\Phi^{T_s}_2(z_\epsilon)), 
\Phi^{T_s}_2(z_\epsilon)\Big )\in W^l_s
$$
the parallel projection of $\Phi^{T_s}(z_\epsilon)$ on the local stable manifold
(see figure \ref{deltas}), that is the point on $W^l_s$ with 
$z_2=\Phi^{T_s}_2(z_\epsilon)$, and by
$$
\Delta_\epsilon = \norm{\Phi^{T_s}_1(z_\epsilon)- w_s(\Phi^{T_s}_2(z_\epsilon)) }  
$$
the distance between $\Phi^{T_s}(z_\epsilon)$ and the point $\pi_\epsilon$. 
Since $\Delta_\epsilon$  depends continuously on $\epsilon$, 
$\Delta_0=0$, and the local stable manifold is invariant, there exists 
$\epsilon_1$ such that $\Delta_\epsilon$ is strictly monotone increasing 
function of $\epsilon\in [0,\epsilon_1]$. We have also (see Section 
\ref{proofs}):
\begin{equation}
\Delta_\epsilon  \leq (1+\lambda_w)\lambda_{\Phi}^{T_s}\epsilon  ,
\label{deltaeps}
\end{equation}
so that if $(1+\lambda_w)\lambda^{T_s}\epsilon < \delta^2$ we have
$\pi_\epsilon \in B(\delta)$. We use $\Delta_\epsilon$ to parameterize the distance 
of $z_\epsilon$ from  the stable manifold $W_s$, and we introduce
the time
\begin{equation}
T_\epsilon = \left [ {1\over\ln \lambda_u} {\ln {e\delta\over \Delta_\epsilon}}\right ]
\label{teps}
\end{equation}
which, as we will prove (see Lemma \ref{lemma01}), is required by the orbit 
with initial condition $\Phi^{T_s}(z_\epsilon)$ to exit from $B(\delta)$.
We also denote by
$$
\zeta_\epsilon =\Big ( \Phi^{T_s+T_\epsilon}_1(z_\epsilon),
w_u(\Phi^{T_s+T_\epsilon}_1(z_\epsilon) )\Big ) \in W^l_u
$$
the parallel projection of $\Phi^{T_s+T_\epsilon}_1(z_\epsilon)$ over the 
local unstable manifold. 
\begin{proposition}\label{pro1}
Let us consider any large $T$ satisfying
\begin{eqnarray}
e\delta \lambda_u^{-\alpha (T-T_s)} &\leq& \Delta_{\epsilon_1} \label{conteps1}\\
e\lambda_u^{-\alpha (T-T_s)} &\leq & {\sigma^{T_s}\delta_0\over  
\max(1,\eta) (1+\lambda_w)\lambda^{T_s}T^2}\label{conteps2}\\
T &>& T_s +{1\over 1-\alpha}\label{conteps3}
\end{eqnarray}
with 
$$
\alpha =  {\ln \lambda \over \ln \lambda+ \ln \lambda_u}  .
$$
By denoting with $\epsilon_0$ the constant such that
\begin{equation}
\Delta_{\epsilon_0} = e\delta \lambda_u^{-\alpha (T-T_s)}   ,
\label{delepszero}
\end{equation}
then, for any $\epsilon \leq \epsilon_0$, if $T_\epsilon \geq T-T_s$ we have
\begin{equation}
\Norm{D\Phi^{T}_{z_\epsilon}v  -A^{T-T_s}w}\leq\lambda_u^{T-T_s} {\Norm{w_u}\over T}\ \ ,\ \ w=D\Phi^{T_s}_{z_s}v  ,
\label{vtanvicino}
\end{equation}
if  $\alpha( T -T_s)\leq  T_\epsilon < T -T_s$  we have
\begin{equation}
{\Norm{D\Phi^{T}_{z_\epsilon}v} \over 
\Norm{D\Phi^{T}_{z_s}v}}\leq \left ( 1+{1\over T}\right )
{\Norm{D\Phi^{j}_{\zeta_\epsilon}}\over \lambda_u^j}\ \ ,\ \ j=T-T_s-T_\epsilon  .
\label{flilontano}
\end{equation}
\end{proposition}
\vskip 0.4 cm
\noindent
The proof is reported in Section \ref{proofs}. 
\vskip 0.4 cm
\noindent
{\bf Remark.}  Conditions (\ref{conteps1}), (\ref{conteps2}) and 
(\ref{conteps3}) may be all satisfied by times $T$ which are suitably large, 
but not necessarily extremely large, because of the presence of the exponentials
in (\ref{conteps1}) and (\ref{conteps2}), and because of the typical dependence
$T_s(\delta)\sim \ln (1/\delta) \sim \ln T$. Therefore, the proposition
is meaningful also for $\epsilon_0$ which are small, but not necessarily 
extremely small. Moreover, from the definition of $\epsilon_0$, apart from a 
small difference due to the use of the integer part in the definition of 
$T_\epsilon$, we have $T_{\epsilon_0} \sim \alpha (T-T_s)$, and 
$T-T_s-T_\epsilon \leq T_u := (T-T_s)(1-\alpha)$. \hfill $\Box$
\vskip 0.4 cm
For $z_s\in W_s$, and for all the points $z_\epsilon$ which are so close to the 
stable manifold that $T_\epsilon \geq T-T_s$, the FLI is approximated by 
$$
\ln\Norm{A^{T-T_s}w}=(T-T_s)\ln \lambda_u+\ln\Norm{w_u}  .
$$ 
Therefore, the only possibility for the FLI to strongly decrease by 
increasing $\epsilon$ is that, for  $\alpha (T-T_s) \leq  T_\epsilon < T-T_s$, 
we have an exponential decrement of $\Norm{D\Phi^{j}_{\zeta_\epsilon}}/\lambda_u^j$ 
with respect to $j$. The assumption which guarantees a 
desired scaling of the FLI with respect 
to $\epsilon$  is 
\begin{equation}
\sup_{\epsilon: \alpha(T-T_s)\leq T_\epsilon \leq T-T_s}
 { \Norm{D\Phi^{T-T_s-T_\epsilon}_{\zeta_\epsilon}}\over (C\lambda_u)^{T-T_s-T_\epsilon}} 
\leq 1
\label{condition}
\end{equation} 
with some $C<1$, so that we have
$$
\ln \Norm{D\Phi^{T}_{z_\epsilon}v}  \leq \ln \Norm{D\Phi^{T}_{z_s}v} 
-(T-T_s-T_\epsilon) \norm{\ln C}+\ln \left ( 1+{1\over T}\right )  .
$$
From the definition of $T_\epsilon$, we have therefore a linear
decrement of the FLI with respect to $\ln \Delta_\epsilon$, up to 
the maximum value of $T-T_s-T_\epsilon\leq (1-\alpha)(T-T_s)$. Therefore, 
at the exponentially small distance from the manifold (\ref{delepszero}) 
the FLI has decreased of a quantity which is proportional 
to integration time $T$, and conversely, the differences of units in the 
FLI value typically determines a proportional number of precision digits in  
the localization of the stable manifold. 

With evidence, condition (\ref{condition}) may be satisfied if 
$\Norm{D\Phi_{z}}$ has an absolute maximum for 
$z\in \cup_{k \leq T_u} \Phi^{-k}(W_u^l)$. For example, 
the condition may be satisfied for the map (\ref{modelmap}) with
$\sigma=-1$, since the origin is a local strict maximum for $\Norm{D\Phi_z}$, 
$z\in W_u$, while it is not satisfied for $\sigma=1$,
since in this case the origin is a local strict minimum for $\Norm{D\Phi_z}$, 
$z\in W_u$.  In any case, it is not 
practical to verify if condition (\ref{condition}) is satisfied by a certain choices of the parameters. Therefore, at the light of the above analysis, we consider a 
generalization of the FLI indicators which depend on a function 
$$
u : {\Bbb R}^2\rightarrow {\Bbb R}^+
$$
as follows: let us consider $z\in M$, $v\in {T_zM}$, and $T>0$. Then, 
we consider $l_T(z,v)$ defined as the $T$--th element of the sequence
\begin{equation}
l_{1}=\ln \Norm{v}\ \ ,\ \ 
l_{j+1}=l_j + u({ z_j}) \ln { \Norm{D\Phi_{z_j} v_j}\over \Norm{v_j}} ,
\end{equation}
{ where $z_j:=\Phi^j(z)$ and $v_j:=D\Phi^j_{z_j}v$}.
The usual FLI is obtained by $u(z)=1$ for any $z\in M$. We consider the 
alternative case of functions $u(z)$ which are test functions 
of some neighbourhood ${\cal B}\subseteq M$ of the fixed point, 
 and precisely with
$u(z)=1$ for $z\in {\overline {\cal B}}$, and $u(z)=0$ for $z$ outside a given 
open set $V \supseteq {\overline {\cal B}}$. We remark that 
the set ${\cal B}$ needs to be small, but not necessarily extremely small. 
For example, if  ${\cal B} \subseteq B(\delta)$, we only 
need, in $V \backslash {\cal B}$,
$$
\Norm{D\Phi_z }^{u(z)} \leq C\lambda_u 
$$
for some $C<1$. The function $u$ described above depends on a specific 
hyperbolic fixed point. If one is interested in the stable or unstable 
manifolds of more fixed points (or hyperbolic periodic orbits), with the 
same numerical integration of the variational equations, forward and 
backward in time, one may compute the FLI indicators related to the 
different fixed points without increasing significantly the computational 
time, and use the results to find, for example, homoclinic and heteroclinic 
intersections between the different manifolds. If instead, one is interested
in determining with a single numerical integration the largest number of
manifolds in some finite domain $B$, one can divide the domain $B$ in 
many small sets $ {\cal B}_j$, $j\leq N$, and compute the $N$ indicators
FLI$_j$ adapted to the sets ${\cal B}_j$. This procedure increases the 
computational time only logarithmically with $N$, since the time required for 
the numerical localization of a point in one of the sets ${\cal B}_j$ 
increases logarithmically with $N$. Then, the portrait of all the manifolds 
is obtained by representing, for any initial condition, the maximum between 
all the FLI$_j$.  Therefore, at variance with the traditional FLI indicator, 
the modified indicators are proposed as a general tool for the numerical 
detection of the stable and unstable manifolds.

\section{A numerical example: the tube manifolds of $L_1$ and $L_2$ in the 
planar circular restricted three body problem}\label{Section3}

\begin{figure}
\begin{center}
\includegraphics[height=6cm,width=9cm]{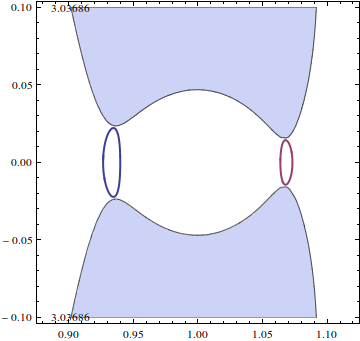}
\end{center}
\caption{Projection on the plane x-y of the Lyapunov orbits related to the
points $L_1$ and $L_2$, for the value  
$C=3.03685733643946038606918461928938$ of the Jacobi constant.
The shaded area represents a region of the orbit plane
which is forbidden for this value of the Jacobi constant.}
\label{L1L2}
\end{figure}

The  circular restricted three-body problem  describes the motion of
a massless body $P$ in the gravitation field of two massive bodies $P_1$
and $P_2$, called primary and secondary body respectively, which 
rotate uniformly around their common center of mass. In a rotating frame $xOy$, 
 the equations of motion of $P$ are:
\begin{equation}\label{threebody}
\left\lbrace \begin{array}{rcl}
\ddot{x}&=&2\dot{y}+x-(1-\mu)\frac{x+\mu}{r_1^3}-\mu\frac{x-1+\mu}{r_2^3}\\
\ddot{y}&=&-2\dot{x}+y-(1-\mu)\frac{y}{r_1^3}-\mu\frac{y}{r_2^3} \\
  \end{array} \right.
\end{equation}
where the units of masses, lengths and time have been chosen so 
that the masses of $P_1$ and $P_2$ are  $1 - \mu$ and $\mu$ ($\mu \le 1/2$)
respectively, their coordinates  are  $(-\mu,0)$ and $(1-\mu,0)$ 
and their revolution period is $2\pi$. We denoted by  $r_1^2=(x+\mu)^2+y^2$ and by $r_2^2=(x-1+\mu)^2+y^2$. As it is well known, equations (\ref{threebody}) 
have an integral of  motion, the so--called Jacobi constant, defined  by:
\begin{equation}\label{jacobi}
{\cal C} (x,y,\dot x, \dot y)=x^2+y^2+2\frac{1-\mu}{r_1}+
2\frac{\mu}{r_2}-\dot{x}^2-\dot{y}^2 ,
\end{equation}
and five equilibria usually denoted by $L_1,\ldots ,L_5$. Here we consider 
$\mu =0.0009537$, which corresponds to the Jupiter--Sun mass ratio value,  
and a value of the Jacobi constant slightly smaller than 
${\cal C}(x_{L_2},0,0,0):=C_2$. As it is extensively explained in 
\cite{KLMR06}, in these conditions, one may find particularly 
interesting dynamics, which we briefly summarize. The equilibrium 
points $L_1,L_2$ are partially hyperbolic, and their center manifolds $W^c_{L_1},
W^c_{L_2}$ are two--dimensional, and foliated near $L_1,L_2$ respectively by 
periodic orbits called Lyapunov orbits. For values $C$ of the Jacobi constant 
slightly smaller than $C_2$, there exist one Lyapunov orbit related 
to $L_1$ and one Lyapunov orbit related to $L_2$ respectively with 
Jacobi constant equal to $C$ (see figure \ref{L1L2}).

 The Lyapunov orbits are 
hyperbolic, and transverse intersections of their stable and unstable 
manifolds--usually called {\it tube manifolds}-- produce the complicate 
dynamics related to the heteroclinic chaos. The numerical computation of the  
 tube manifolds has been afforded in several papers, and has important 
implications also for modern space mission design (see \cite{Simo99}, 
\cite{KLMR06}). 

In this Section we analyze the FLI method for the detection 
of the tube manifolds introduced in \cite{LGF11,GL13mnras} at the light 
of the theoretical analysis performed in Section \ref{Section2}, and 
we show that the method allows for a detection of the manifolds  
with a number of precision digits which increase linearly with 
respect to the integration time. Moreover, the modified FLI allows us  
to compute the manifolds with a precision limited only by 
the round--off of the numerical computations. 

We report here three numerical experiments. 
In the first one we illustrate the numerical precision of the FLI method
in the determination of the stable tube manifold of a Lyapunov 
periodic orbit around $L_1$; in the second one,  we provide some snapshots 
of the stable tube manifold of the Lyapunov 
periodic orbit around $L_2$ and the unstable tube manifold of the Lyapunov 
periodic orbit around $L_1$, obtained by extending the integration time;   
in the third one we illustrate the numerical precision of the FLI method 
for the localization of a heteroclinic 
intersection between these two manifolds. We 
remark that these computations are particularly critical since 
the tube manifolds are located in a region of the phase space
close to the singularity at $(x,y)=(1-\mu,0)$. In these circumstances, 
the numerical computation of both equations of motions (\ref{threebody}) 
and their variational equations becomes critical, and several 
approaches have been introduced (see \cite{Villac08,LGF10b,clsf,GL13mnras}).  

\begin{figure}
\begin{center}
\includegraphics[height=6cm,width=9cm]{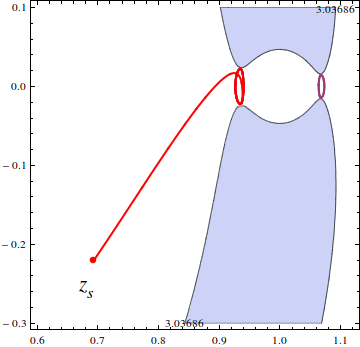}
\end{center}
\caption{ Projection on the plane $(x,y)$ of an orbit with initial condition
{ $z_s=(x_s,y_s,\dot x_s,\dot y_s)\in W^s_{L_1}$, 
with $x^s=0.687020836763335598413507147121355$,
$y^s=-0.227669455733293321520979535995733$, $\dot x^s = 0.331597964276881596512604348842892$, and $\dot y^s$ obtained from the Jacobi constant $C=3.03685733643946038606918461928938$. The shaded area represents a region of the orbit plane
which is forbidden for the value $C$ of the Jacobi constant.}}
\label{stab1}
\end{figure}

\begin{figure}
\vskip -8 truecm
\hskip 2truecm
\includegraphics[height=12cm,width=14cm]{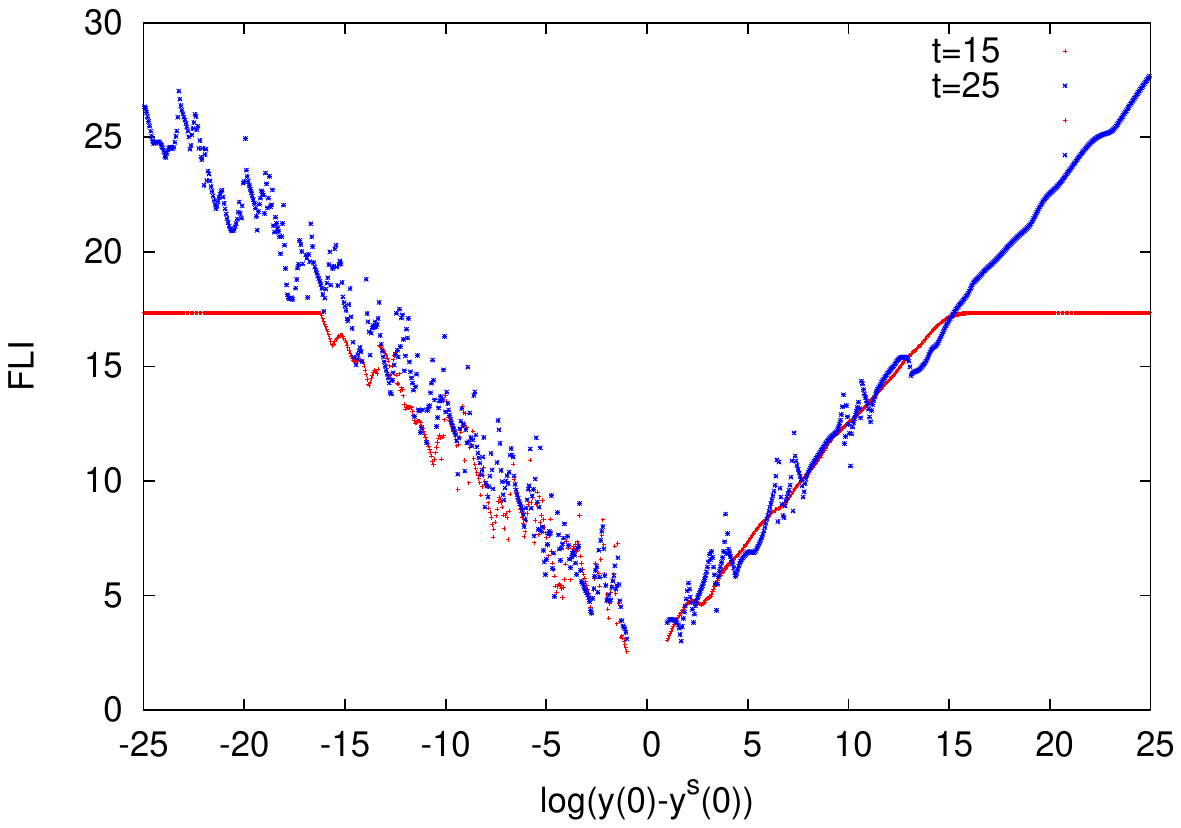}
\caption{Values of the traditional FLI computed on a set of 960 initial
conditions with { $(x(0),y(0))=(x_s,\dot x_s)$ (see Fig.\ref{stab1}), 
$\log\norm{y(0)-y_s}$ in the interval $[-25,-1]$ and $\dot y(0)$ 
obtained from the Jacobi constant $C=3.03685733643946038606918461928938$.
The integration times are respectively $T=15$ and $T=25$, (the negative 
values correspond to initial conditions with $y(0)<y^s$). We appreciate 
a localization of the manifold determined by a linear decrement of the FLI 
with respect to $\log\norm{y(0)-y_s}$. The time $T=15$ allows us to localize
the manifold with a precision of order $10^{-15}$, while the time
$T=25$  allows us to localize the manifold more precisely than  $10^{-22}$. }}
\label{Flitotstab1}
\end{figure}

For the computation of the tube manifolds, we find 
particularly useful to define the variational equation in the space
of the variables obtained by regularizing equations (\ref{threebody})
with respect to the secondary mass, as in \cite{ clsf,GL13mnras}. Precisely, 
we consider the Levi--Civita regularization defined by the 
space transformation 
 \begin{equation}
\left\lbrace \begin{array}{lll}
  x-(1-\mu) & = &u_1^2-u_2^2  \\
 y &=&  2 u_1u_2  \\
 \end{array} \right .
\end{equation}
and by the fictitious time $s$ related to $t$ by $dt = r_2 ds$. The equations 
of motion in the variables $u_1,u_2$, and fictitious time  $s$ are (see for example \cite{Szebe67}):
 \begin{equation}\label{motio1}
\left\lbrace \begin{array}{lll}
u_1'' & = &{1\over 4}[ (a+b)u_1+cu_2]  \\
u_2'' & = &{1\over 4}[ (a-b)u_2+cu_1]  \\
 \end{array} \right .
\end{equation}
with:
 \begin{equation}\label{motio2}
\left\lbrace \begin{array}{lll}
a & = &  {\frac {2(1-\mu)}{r_1}}-C+x^2+y^2 \\
b & = & 4y'+2r_2 x-{\frac {2(1-\mu) r_2(x- 1+\mu)}{r_1^3}} \\
c & = &  2r_2 y-4x'-{\frac {2(1-\mu) r_2 y} {r_1^3}} 
 \end{array} \right .
\end{equation}
where $C$ denotes the value of the Jacobi constant, and the primed 
derivatives denote derivatives with respect the fictitious 
time $s$. To define the FLI, we first write (\ref{motio1}) as a system of 
first order differential equations:
 \begin{equation}
\left\lbrace \begin{array}{lll}
u'_1 &=& v_1\\
u'_2 &=& v_2\\
v_1' & = &{1\over 4}[ (a+b)u_1+cu_2]  \\
v_2' & = &{1\over 4}[ (a-b)u_2+cu_1]  \\
 \end{array} \right .
\end{equation}
and we introduce its compact form:
\begin{equation}
\xi' = F (\xi)\\
\label{compacttb}
\end{equation}
with $\xi=(u_1,u_2,v_1,v_2)$. The variational equations 
of (\ref{compacttb}) are therefore:
\begin{equation}
\left\lbrace
\begin{array}{lcr}
&\xi' = F (\xi)&\cr
&w' = {\partial F\over \partial \xi}(\xi) w&  ,
\end{array} \right .
\label{varsys}
\end{equation}
where $w\in \Bbb R^4$ represents a tangent vector. Following
\cite{GL13mnras}, we here consider the regularized FLI indicator defined by
\begin{equation}
FLI(\xi(0),w(0),T) =  \log  \Norm{w({ s(T)})}
\label{RFLI0}
\end{equation}
where $\xi(s),w(s)$  denotes the solution of the variational equations
(\ref{varsys}) with initial condition $\xi(0),w(0)$ and $s(T)$ is 
the fictitious time which corresponds to the physical time $T$ for that 
orbit. The indicator (\ref{RFLI0}) will be computed also for 
negative times $T<0$. 
\vskip 0.4 cm
\noindent
{\bf FLI detection of the tube manifolds.} In order to test the precision 
of the FLI method in the localization 
of the tube manifolds, we consider a point $z_s=(x_s,y_s,\dot x_s,\dot y_s)
\in W^s_{L_1}$ in the stable tube 
manifold of the Lyapunov orbit around $L_1$ (see Figure \ref{stab1}), and we 
compute the traditional and modified FLIs for a set of many initial conditions. 
with $(x(0),y(0))=(x_s,\dot x_s)$ (see Fig.\ref{stab1}), 
$\log\norm{y(0)-y_s}$ in the interval $[-25,-1]$ and $y(0)$ 
obtained from the value of the Jacobi constant 
$C=3.03685733643946038606918461928938$. The integration times are 
respectively $T=15$ and $T=25$. We appreciate 
a localization of the manifold  determined by a linear decrement of the FLI 
with respect to $\log\norm{y(0)-y_s}$. The time $T=15$ allows us to localize
the manifold with a precision of order $10^{-15}$, which is greatly improved 
by using $T=25$. We obtain a good localization of the manifold already 
with the traditional FLI, see Figure \ref{Flitotstab1}, 
although the irregularities in the FLI curve limit the precision of the 
localization to $10^{-22}$, higher than the numerical round--off precision. 

Then, we considered 
a  modified FLI defined by equations
(\ref{ltind}) with function $u(z)$ which is a test function of a neighbourhood 
 of the Lyapunov orbit $\gamma_1$ around $L_1$. Precisely, we use a test 
function defined by:
\begin{equation}
u(z)=\left \lbrace \begin{array}{lcr}
&1&\ {\rm if}\ \ \norm{z-\gamma_1}\leq {r_1 \over 2} \\
& {1\over 2}[{\cos (({\norm{z-\gamma_1}\over r_1}-{1\over 2})\pi )+1}]  &  {\rm if}\ \ {r_1 \over 2} < \norm{z-\gamma_1}\leq  {3r_1\over 2}\\
&0& \ {\rm if}\ \ \norm{z-\gamma_1}> {3 r_1 \over 2}
\end{array}\right .  
\end{equation}
where $\norm{z-\gamma_1}$ denotes the distance between $z$ and the 
Lyapunov orbit $\gamma_1$ (we set $r_1=10^{-3}$ in the following computations).
 Also in this case the time $T=15$ allows us to 
localize
the manifold with a precision of order $10^{-15}$, while the time
$T=25$  allows us to localize the manifold more precisely than  $10^{-25}$. 
The use of the modified FLI has eliminated the irregularities in 
the curves of Figure \ref{Flitotstab1}, and improved the precision of the 
localization. As a matter of fact, the precision of the localization is 
reduced to the round--off used for the numerical computation.

\begin{figure}
\vskip -8 truecm
\hskip 2truecm
\includegraphics[height=12cm,width=14cm]{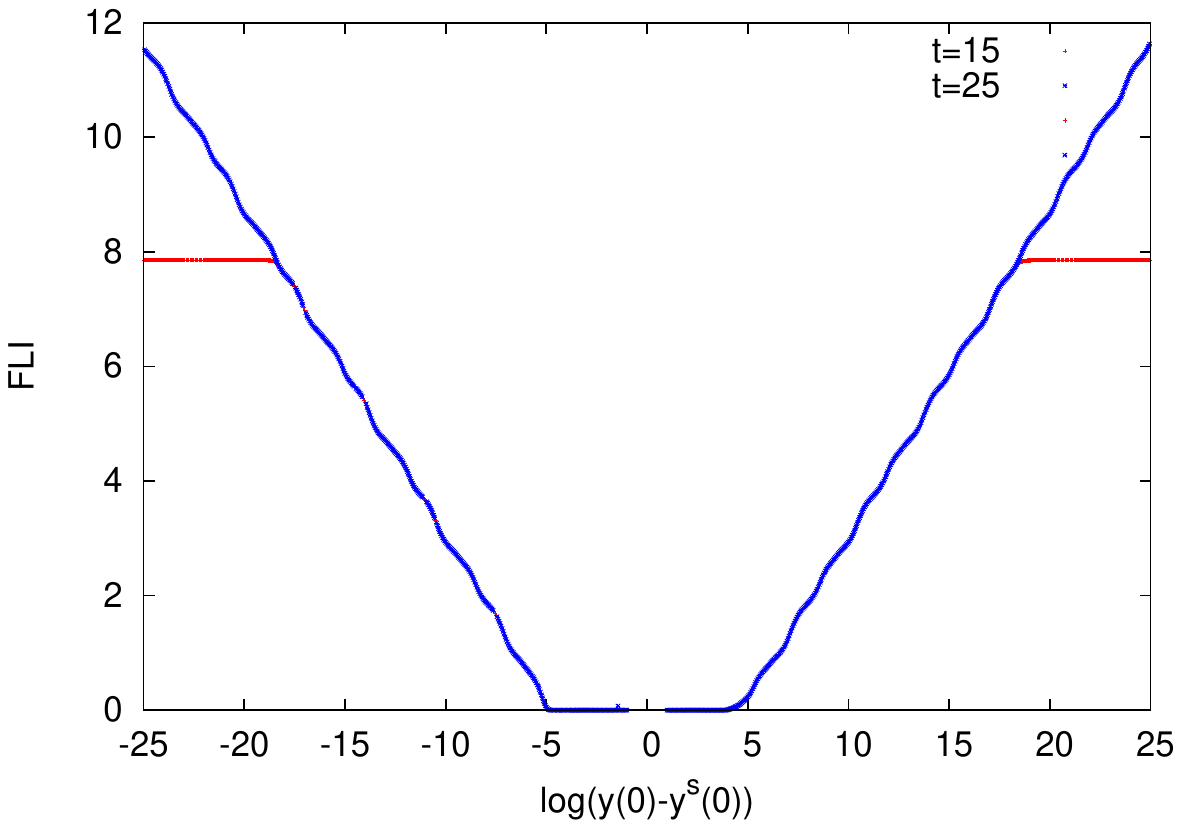}
\caption{{ Values of the modified FLI defined by equations
(\ref{ltind}) with function $u(z)$ which is a test function of a neighbourhood 
 of the Lyapunov orbit around $L_1$. The initial conditions are the 
same 960 initial conditions 
considered in Figure \ref{Flitotstab1}, that is  
$(x(0),y(0))=(x_s,\dot x_s)$ (see Fig.\ref{stab1}), 
$\log\norm{y(0)-y_s}$ in the interval $[-25,-1]$ and $y(0)$ 
obtained from the Jacobi constant $C=3.03685733643946038606918461928938$.
The integration times are respectively $T=15$ and $T=25$, (the negative 
values correspond to initial conditions with $y(0)<y_s$). We appreciate 
a localization of the manifold determined by a linear decrement of the FLI 
with respect to $\log\norm{y(0)-y_s}$. The time $T=15$ allows us to localize
the manifold with a precision of order $10^{-15}$, while the time
$T=25$  allows us to localize the manifold more precisely than  $10^{-25}$. 
The use of the modified FLI has eliminated the irregularities in 
the curves of Figure \ref{Flitotstab1}, and improved he precision of the 
localization. As a matter of fact, the precision of the localization is 
reduced to the round--off used for the numerical computation.   
}}
\label{Fliboxstab1}
\end{figure}

\begin{figure}
\vskip -1truecm
\includegraphics[height=9cm,width=11cm]{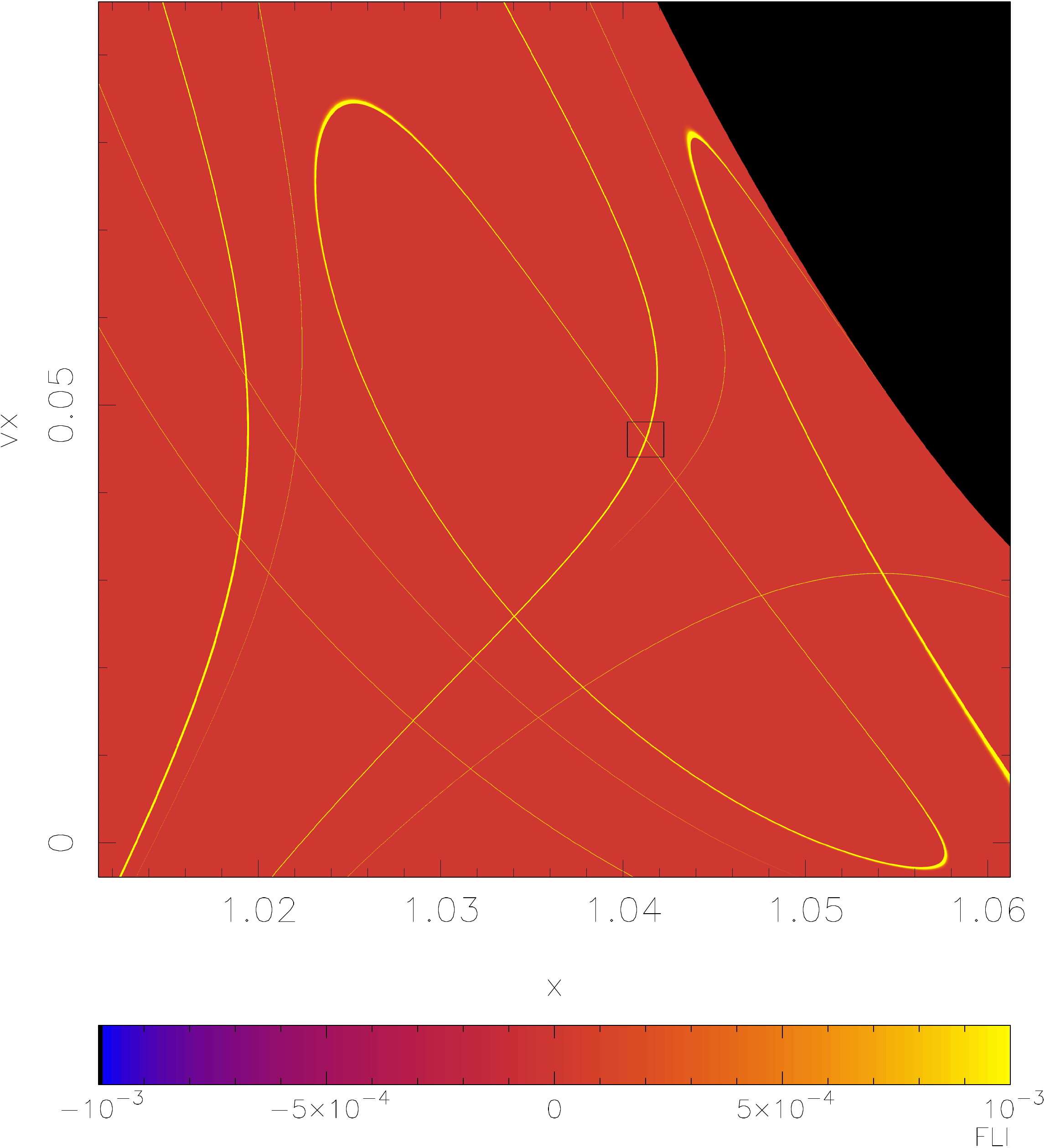}
\caption{Representation of the modified FLIs computed on a grid of $4000\times 4000$ 
initial conditions regularly spaced on $(x,\dot x)$ (the axes 
on the picture--the other initial conditions
are  $y=0$ and $\dot y$ is computed from the Jacobi constant 
$C=3.0368573364394607$), computed with integration time $T=5$.
In order to represent both manifolds on the same 
picture, we represent with a color scale the weighted average (\ref{wav}) 
of the two indicators $ {\rm FLI}_1$, ${\rm FLI}_2$ with weight $w=100$. 
The yellow curves on the picture correspond to different lobes of the manifolds.
}
\label{tubemanif5}
\end{figure}

\begin{figure}
\vskip -2truecm
\includegraphics[height=9cm,width=11cm]{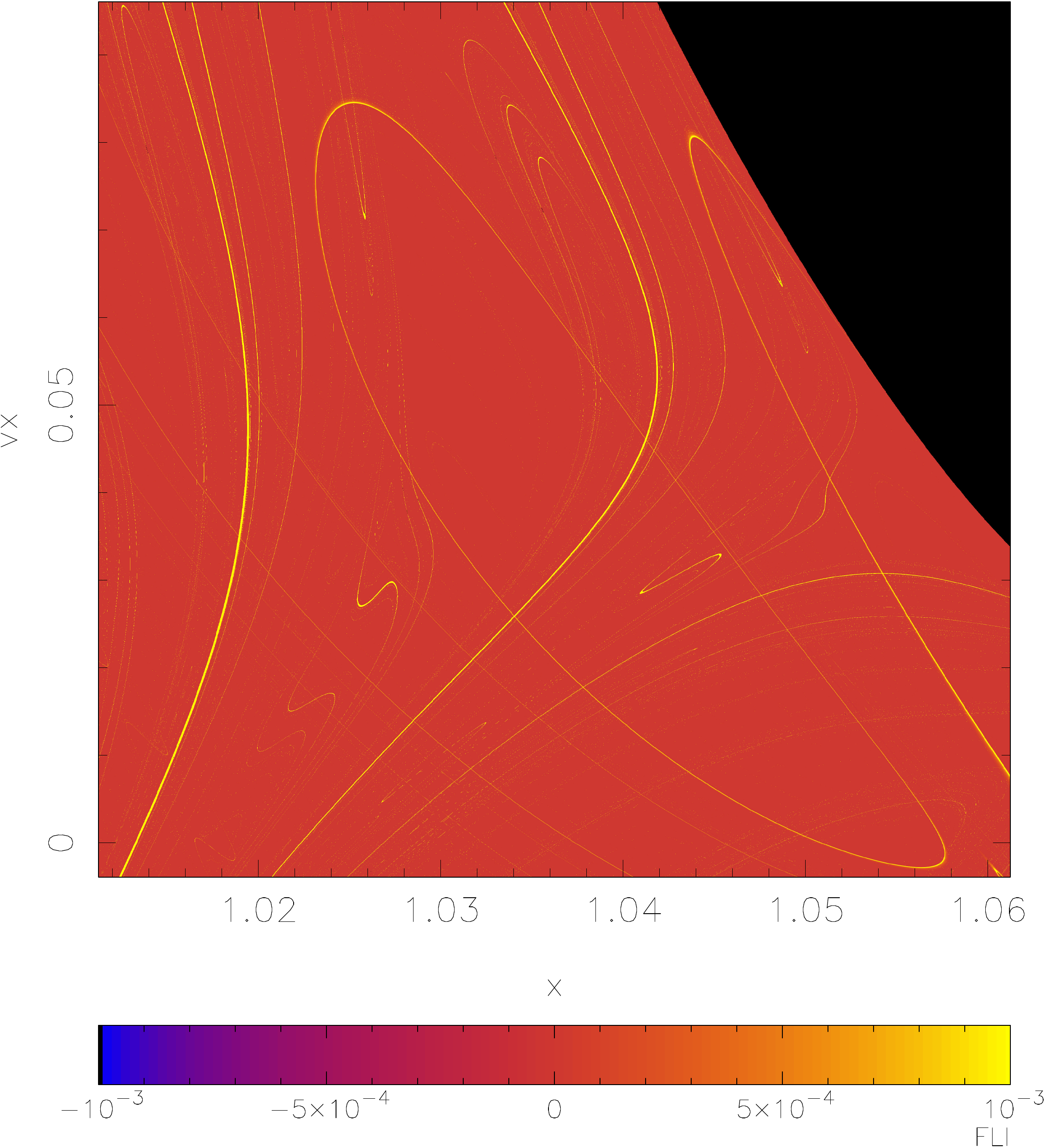}
\caption{Representation of the modified FLIs computed on a grid of 
$4000\times 4000$ 
initial conditions regularly spaced on $(x,\dot x)$ (the axes 
on the picture--the other initial conditions
are  $y=0$ and $\dot y$ is computed from the Jacobi constant 
$C=3.0368573364394607$), computed with an integration time $T=100$. 
In order to represent both manifolds on the same 
picture, we represent with a color scale the weighted average (\ref{wav}) 
of the two indicators $ {\rm FLI}_1$, ${\rm FLI}_2$ with weight $w=500$. 
The yellow curves on the picture correspond to different lobes of the manifolds.
Due to the integration time which is much longer than the time 
used in Figure \ref{tubemanif5}, many additional lobes of the tube 
manifolds of both $\gamma_1$ and $\gamma_2$ appear on this figure. 
Their corresponding initial conditions approach the manifolds only 
after several revolution periods of Jupiter. }
\label{tubemanif100}
\end{figure}

\vskip 0.4 cm
\noindent
{\bf Snapshots of tube manifolds of $W^u_{L_1}$ and $W^s_{L_2}$.} Motivated by these  
results, we obtained sharp representations of the intersections
$$
W^s_{L_2} \cap \Sigma\ \ ,\ \ W^u_{L_1} \cap \Sigma
$$
of the stable tube manifold $W^s_{L_2}$ of the 
Lyapunov orbit $\gamma_2$ around $L_2$ and of the unstable tube 
manifold $W^u_{L_1}$ of the Lyapunov orbit $\gamma_1$ around $L_1$ with the 
two--dimensional section of the phase--space defined by
\begin{equation}
\Sigma = \{ (x,y,\dot x,\dot y): \ \ y=0\ \ ,\ \ \dot y\geq 0:\ \ 
{\cal C}(x,0,\dot x, \dot y)=C\}  .
\label{section} 
\end{equation}
Any point $z\in \Sigma$ is parameterized and identified 
by its two components $(x,\dot x)$. The representation of the manifolds are obtained 
by computing the modified FLIs on  refined grids of initial 
conditions $(x,\dot x)$ on $\Sigma$ for different integration
times $T$. The stable manifold  $W^s_{L_2}$ is obtained by computing 
the modified FLI on a time $T_2$, using a test function defined by 
\begin{equation}
u(z)=\left \lbrace \begin{array}{lcr}
&1&\ {\rm if}\ \ \norm{z-\gamma_2}\leq {r_2 \over 2} \\
& {1\over 2}[{\cos (({\norm{z-\gamma_2}\over r_2}-{1\over 2})\pi )+1}]  &  {\rm if}\ \ {r_2 \over 2} < \norm{z-\gamma_2}\leq  {3r_2\over 2}\\
&0& \ {\rm if}\ \ \norm{z-\gamma_2}> {3 r_2 \over 2}
\end{array}\right .  
\label{test2}
\end{equation}
where $\norm{z-\gamma_2}$ denotes the distance between $z$ and the 
Lyapunov orbit $\gamma_2$ and $r_2=5\,10^{-4}$. The unstable manifold  
$W^u_{L_1}$ is obtained by computing the modified 
FLI on a negative time $-T_1$, using a test function 
defined by 
\begin{equation}
u(z)=\left \lbrace \begin{array}{lcr}
&1&\ {\rm if}\ \ \norm{z-\gamma_1}\leq {r_1 \over 2} \\
& {1\over 2}[{\cos (({\norm{z-\gamma_1}\over r_1}-{1\over 2})\pi )+1}]  &  {\rm if}\ \ {r_1 \over 2} < \norm{z-\gamma_1}\leq  {3r_1\over 2}\\
&0& \ {\rm if}\ \ \norm{z-\gamma_1}> {3 r_1 \over 2}
\end{array}\right .  
\label{test1}
\end{equation}
where $\norm{z-\gamma_1}$ denotes the distance between $z$ and the 
Lyapunov orbit $\gamma_1$ and $r_1=10^{-3}$. In such a way, for any $x,\dot x$ 
we compute the modified FLIs: FLI$_1$, FLI$_2$. The representation of
both manifolds on the same picture is obtained by representing with 
a color scale a weighted average of the two indicators: 
\begin{equation}
{w {\rm FLI}_1+{\rm FLI}_2 \over (w+1)}  .
\label{wav}
\end{equation}
The results are represented in Figures \ref{tubemanif5} and 
\ref{tubemanif100} for $T=5$ and $T=100$ respectively. 
We clearly appreciate different lobes of both manifolds already 
for the shorter integration time   $T=5$. 
The longer time $T=100$ allows us to appreciate additional lobes, which 
contain initial condition approaching the manifolds only after several 
revolution periods of Jupiter. 


\vskip 0.4 cm
\noindent
{\bf Localization of heteroclinic intersections.} The detection  of 
both manifolds $W^u_{L_1}$ and $W^s_{L_2}$ on the same picture (see Figure 
\ref{tubemanif5} and Figure \ref{tubemanif100}) allows us to 
obtain a precise localization of the heteroclinic intersections points, 
which we denote by $z_{he}$. Precisely,  the intersection between the two 
yellow curves in the  box of Fig.\ref{tubemanif5} corresponds to an 
intersection point $z_{he}$. Of course, accordingly to the 
resolution of the computation, at first we are only able to determine 
a point $z_{he,1}$ in the box which is close $z_{he}$. To improve
the localization of $z_{he}$ we compute again the modified FLIs on a 
refined grid of  points in the box of Fig.\ref{tubemanif5}, and we
obtain a new point $z_{he,2}$ (the point with the maximum value 
of the averaged FLI (\ref{wav})) closer to  the intersection point.
The procedure is iterated by computing again the FLIs on zoomed out
 grids of initial conditions centered on $z_{he,j}$ with $j=2,...15$, 
with increasing integration times to increase the number of precision digits 
in the localization of the heteroclinic point. \par
\begin{figure}
\vskip -1truecm
\begin{center}
\includegraphics[height=7cm,width=10cm]{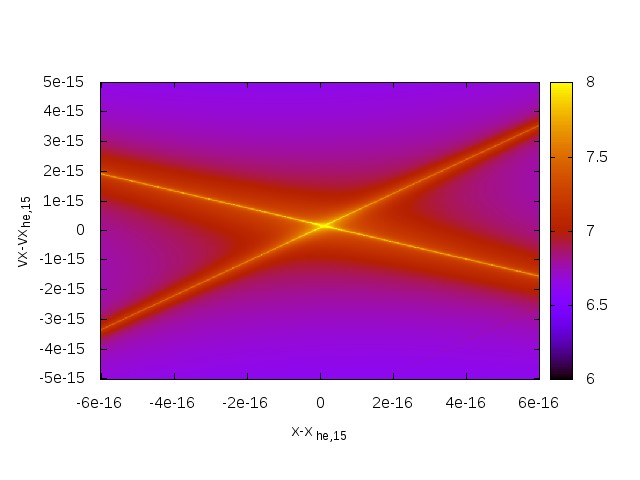}
\end{center}
\caption{Computation of the averaged FLI (\ref{wav}) on a grid of 
$500\times 500$  initial conditions centered on the point $z_{he,15}$ of 
coordinates:   $x_{he,15}=1.041239777351473900$,
 $y_{he,15}=0$, $ \dot x_{he,15}=0.0460865533656582000$. The velocity 
$\dot y_{he,15}$ is obtained from the Jacobi constant $C=3.0368573364394607$.  The integration time is $T=18$.
The values of the FLI are provided as  the average  between  FLI$_1$
and  FLI$_2$. A sharp  detection of both manifolds appears thanks to the differentiation of the FLI values on this refined grid. The maximum value of the FLI in this picture provides a new refined initial condition for the orbit plotted in Fig.\ref{heter}. }
\label{zoom15}
\end{figure}

\begin{figure}
\begin{center}
\includegraphics[height=7cm,width=10cm]{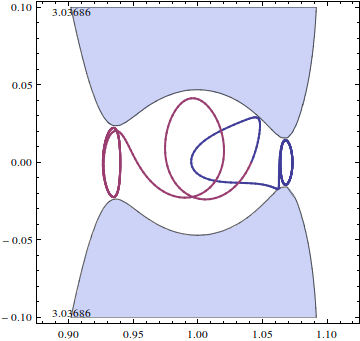}
\end{center}
\caption{ Projection on the plane $(x,y)$ of the heteroclinic orbit found 
through the maximum of the FLI (see text). The 
conditions are : $x_{he}(0)=1.041239777351473912$,
$y_{he}(0)=0$, $\dot x_{he}(0)=0.046086553365658360$ and $\dot y_{he}(0)$ obtained from the Jacobi constant $C=3.0368573364394607$. Blue points: forward integration,
the orbit converges to the Lyapunov orbit related to $L_2$. Red points: backward  integration, the orbit converges to the Lyapunov orbit related to $L_1$. }
\label{heter}
\end{figure}

 
In Fig.\ref{zoom15} we plot the FLI values computed on a grid of 
$500\times 500$  initial conditions centered on the point $z_{he,15}$, 
using the integration time $T=18$. 
The maximum value of the FLI in this picture provides a new refined initial condition that we used to compute the heteroclinic orbit shown in Fig.\ref{heter}.
The convergence of the forward (backward) integration towards   the Lyapunov orbit related to $L_2$ ($L_1$) clearly shows the validity of the method for the precise localization of   heteroclinic orbits. 


\section{Proofs}\label{proofs}

\noindent
{\bf Proof of Proposition \ref{pro1}.} We first remark that 
(\ref{conteps1}) implies $\epsilon_0\leq \epsilon_1$, and condition 
(\ref{conteps2}) implies
$$
\epsilon_0 \leq {\delta_0^2\over \max(1,\eta) (1+\lambda_w)\lambda^{T_s}T^2}  .
$$
The proof of Proposition \ref{pro1} is 
a consequence of the following:
\begin{lemma}\label{lemma01}
For any $\epsilon,\delta$ satisfying
\begin{eqnarray}
&\max(1,\eta) (1+\lambda_w)\lambda^{T_s}\epsilon 
\leq \delta^2&\label{epslem}  \\
&\delta \leq \min \left ( {1\over 2} \ ,\  {r_*\over 2}\  ,\ {1\over \eta}\ ,\ 
{1\over 4e^2\lambda_u\eta}\Big (1-{1\over \lambda_u}\Big )\ ,\ 
{1\over 2e^3\lambda_u^2\eta}{1\over T_\epsilon}\right )&\label{deltlem}
\end{eqnarray}
we have:
\begin{equation}
\delta - \eta \Delta_\epsilon \leq    
\norm{  \Phi^{T_s+T_\epsilon}_1(z_\epsilon)} \leq 
 e^2 \lambda_u \delta  +  \eta \Delta_\epsilon 
\end{equation}
\begin{eqnarray}
&\norm{\Phi^{T_s+T_\epsilon}_2(z_\epsilon)} \leq e^2 \lambda_u \delta  +  
2\eta \Delta_\epsilon &\\
&\norm{\Phi^{T_s+T_\epsilon}_2(z_\epsilon) - (\zeta_\epsilon)_2} \leq 
{1\over \lambda_u^{T_\epsilon}}4 e^3\lambda_u\delta&  ,
\label{diffwl}  
\end{eqnarray}
and the tangent vector $D\Phi^{T_s+T_\epsilon}_{z_\epsilon}v$ satisfies
\begin{equation}
\Norm{D\Phi^{T_s+T_\epsilon}_{z_\epsilon}v  -A^{T_\epsilon}(w_u+w_s)}\leq\lambda_u^{T_\epsilon}
\delta \left (\eta  \max(e,\eta){4e^2\lambda_u^2 \over  \lambda_u-1}+1
\right )\Norm{w_u}  .
\label{vtanA2}
\end{equation}
\end{lemma}
First, as we anticipated in Section \ref{Section2}, the time $T_\epsilon$ can be 
identified as the time required 
by the orbit with initial condition $\Phi^{T_s}(z_\epsilon)$ to 
exit from $B(\delta)$ and to arrive at the small distance 
$(4 e^3\lambda_u\delta)/\lambda_u^{T_\epsilon}$ from 
the local unstable manifold. 
\vskip 0.2 cm
\noindent
If  $T_\epsilon \geq T -T_s$, we can repeat the proof of Lemma
\ref{lemma01} by limiting all the estimates to the time interval 
$[0,T]$, and obtaining
\begin{equation}
\Norm{D\Phi^{T}_{z_\epsilon}v  -A^{T-T_s}w}\leq\lambda_u^{T-T_s}
\delta \left (\eta  \max(e,\eta){4e^2\lambda_u^2 \over  \lambda_u-1}+1
\right )\Norm{w_u}  ,
\label{vicino}
\end{equation}
so that (\ref{vtanvicino}) is proved.  
\vskip 0.2 cm
\noindent
If $T_\epsilon < T-T_s$,  we need an estimate of the growth of the 
tangent vectors in the remaining time interval $[T_s+T_\epsilon,T]$, and we 
obtain it by comparison with the growth of the tangent vectors of the orbits 
with initial condition in the point $\zeta_\epsilon$ on the unstable manifold. 
We will provide estimates of the FLI for $T_\epsilon$ in the 
interval:
\begin{equation}
 (T-T_s(\delta)) {\ln \lambda \over \ln \lambda +\ln \lambda_u } 
\leq T_\epsilon < T-T_s(\delta)  .
\label{tinterval}
\end{equation}
Let us consider $j=T-T_s-T_\epsilon \in \{1, (T-T_s(\delta)) 
{\ln \lambda_u \over \ln \lambda +\ln \lambda_u }\}$. First, we have 
\begin{equation}
\Norm{D \Phi^{T}_{z_\epsilon}v-D\Phi^{j}_{\zeta_\epsilon}D\Phi^{T_s+T_\epsilon}_{z_\epsilon}v}\leq 
4e^3 \lambda_u {\lambda^j \over \lambda_u^{T_\epsilon}}\delta
\Norm{ D\Phi^{j}_{\zeta_\epsilon}D\Phi^{T_s+T_\epsilon}_{z_\epsilon}v}  ,
\label{vtanA4}
\end{equation}
In fact, since
$$
D \Phi^{T}_{z_\epsilon}v= D\Phi^{j}_{\Phi^{T_s+T_\epsilon}(z_\epsilon)}D\Phi^{T_s+T_\epsilon}_{z_\epsilon}v= D\Phi^{j}_{\zeta_\epsilon}D\Phi^{T_s+T_\epsilon}_{z_\epsilon}v+
\Big (D\Phi^{j}_{\Phi^{T_s+T_\epsilon}(z_\epsilon)}-D\Phi^{j}_{\zeta_\epsilon}\Big )
D\Phi^{T_s+T_\epsilon}_{z_\epsilon}v  ,
$$
using Lemmas \ref{lemma01} and \ref{lemma2} we obtain
$$
\Norm{D \Phi^{T}_{z_\epsilon}v-D\Phi^{j}_{\zeta_\epsilon}D\Phi^{T_s+T_\epsilon}_{z_\epsilon}v}\leq 
\lambda^j \Norm{\Phi^{T_s+T_\epsilon}_{z_\epsilon}-\zeta_\epsilon}\Norm{ D\Phi^{j}_{\zeta_\epsilon}D\Phi^{T_s+T_\epsilon}_{z_\epsilon}v} 
\leq 4e^3 \lambda_u {\lambda^j \over \lambda_u^{T_\epsilon}}\delta
\Norm{ D\Phi^{j}_{\zeta_\epsilon}D\Phi^{T_s+T_\epsilon}_{z_\epsilon}v}  .
$$
Therefore, we have
$$
\Norm{D \Phi^{T}_{z_\epsilon}v}\leq \Norm{D\Phi^{j}_{\zeta_\epsilon}D\Phi^{T_s+T_\epsilon}_{z_\epsilon}v}\left (1+4e^3 \lambda_u {\lambda^j \over \lambda_u^{T_\epsilon}}\delta \right )
 .
$$
We now analyze and compare the FLI for initial conditions 
at different distances from the stable manifold. We have
$$
 {\Norm{D\Phi^{T}_{z_\epsilon}v} \over 
\Norm{D\Phi^{T}_{z_s}v}}\leq {\Norm{D\Phi^{j}_{\zeta_\epsilon}D\Phi^{T_s+T_\epsilon}_{z_\epsilon}v} \over 
\Norm{D\Phi^{T}_{z_s}v}}
\left (1+4e^3 \lambda_u {\lambda^j \over \lambda_u^{T_\epsilon}}\delta \right )  .
$$ 
Using inequalities (\ref{vicino}) and (\ref{vtanA2}), we obtain 
\begin{equation}
{\Norm{D\Phi^{T}_{z_\epsilon}v} \over 
\Norm{D\Phi^{T}_{z_s}v}}\leq {\Norm{D\Phi^{j}_{\zeta_\epsilon}}\over \lambda_u^j}
\ {1+\delta \left ( \eta\max (e,\eta ){4e^2\lambda_u^2\over 
\lambda_u -1}+1\right )\over 
1-\delta \left (\eta  \max(e,\eta){4e^2\lambda_u^2 \over  \lambda_u-1}+1
\right )}\left ( 1+   4e^3 \lambda_u {\lambda^j \over \lambda_u^{T_\epsilon}}\delta
\right ) .
\label{flicomparis0}
\end{equation}
In fact, from (\ref{vicino}) and $\Norm{A^{T-T_s}(w_u+w_s)} = 
\lambda_u^{T-T_s}\Norm{w_u}$, we obtain that 
for all $\epsilon$ with $T_\epsilon \geq T-T_s$, including $z_0=z_s$, we have 
\begin{equation}
\Norm{D\Phi^{T}_{z_\epsilon}v}\geq \lambda_u^{T-T_s} 
\left (1-\delta \left (\eta  \max(e,\eta){4e^2\lambda_u^2 \over  \lambda_u-1}+1
\right )\right )\Norm{w_u}  .
\label{flilb}
\end{equation}
From (\ref{vtanA2}), for all $\epsilon$ with $T_\epsilon \leq T-T_s$,
 we have:
$$
\Norm{D\Phi^{j}_{\zeta_\epsilon}D\Phi^{T_s+T_\epsilon}_{z_\epsilon}v} \leq 
\Norm{D\Phi^{j}_{\zeta_\epsilon}} \Norm{D\Phi^{T_s+T_\epsilon }_{z_\epsilon}v}
$$
$$
\leq \Norm{D\Phi^{j}_{\zeta_\epsilon}}\left( \Norm{A^{T_\epsilon }w}+
\lambda_u^{T_\epsilon}\delta \left ( \eta\max (e,\eta ){4e^2\lambda_u^2\over 
\lambda_u -1}+1\right )\Norm{w}\right )
$$
\begin{equation}
\leq \Norm{D\Phi^{j}_{\zeta_\epsilon}}\lambda_u^{T_\epsilon}\Norm{w_u}
\left(1+
\delta \left ( \eta\max (e,\eta ){4e^2\lambda_u^2\over 
\lambda_u -1}+1\right )\right )  .
\label{fliub}
\end{equation}
Since for all $\epsilon$ with
$$
T_\epsilon \geq (T-T_s) {\ln \lambda \over \ln \lambda +\ln \lambda_u }  ,
$$
we have 
$$
{\lambda^j \over \lambda_u^{T_\epsilon}}= 
{\lambda^{T-T_s-T_\epsilon} \over \lambda_u^{T_\epsilon}} \leq 1  ,
$$
using also
$$
\delta \left ( \eta\max (e,\eta ){4e^2\lambda_u^2\over 
\lambda_u -1}+1\right ) \leq {1\over 2e T}\ , \  
4e^3 \lambda_u \delta < {1\over 4T}
$$
we have
$$
{1+\delta \left ( \eta\max (e,\eta ){4e^2\lambda_u^2\over 
\lambda_u -1}+1\right )\over 
1-\delta \left (\eta  \max(e,\eta){4e^2\lambda_u^2 \over  \lambda_u-1}+1
\right )}\left ( 1+   4e^3 \lambda_u {\lambda^j \over \lambda_u^{T_\epsilon}}\delta
\right )\leq \left (1 + {1\over T}\right )  ,
$$ 
so that, from (\ref{flicomparis0}), we immediately obtain 
(\ref{flilontano}).

\vskip 0.4 cm
\noindent
{\bf Proof of (\ref{deltaeps}).} We have:
$$
\Delta_\epsilon = \norm{\Phi^{T_s}_1(z_\epsilon)- w_s(\Phi^{T_s}_2(z_\epsilon)) }\leq
\norm{\Phi^{T_s}_1(z_\epsilon)-\Phi^{T_s}_1(z_s)}+
\norm{w_s(\Phi^{T_s}_2(z_s))-w_s(\Phi^{T_s}_2(z_\epsilon))}
$$
$$
\leq \lambda_{\Phi}^{T_s}\epsilon +\lambda_w  
\lambda_{\Phi}^{T_s}\epsilon\leq (1+\lambda_w)\lambda_{\Phi}^{T_s}\epsilon  .
$$

\noindent
{\bf Proof of Lemma \ref{lemma01}.} We consider the segments 
which join $\Phi^k(\pi_\epsilon)$ and $\Phi^k(\Phi^{T_s}(z_\epsilon))$, 
and define
$$
\Delta^k_1 = \norm{\Phi^k_1(\Phi^{T_s}(z_\epsilon))-\Phi^k_1(\pi_\epsilon)}
$$
$$
\Delta^k_2 = \norm{\Phi^k_2(\Phi^{T_s}(z_\epsilon))-\Phi^k_2(\pi_\epsilon)}  .
$$
We prove that, for all the $k$ such that 
$\Phi^k(\pi_\epsilon),\Phi^k(\Phi^{T_s}(z_\epsilon))
\leq B(A\delta)$, for $A> 1$, we have
$$
\Delta^k_2 < \Delta^k_1  .
$$
In fact, we have $\Delta^0_1=\Delta_\epsilon$, $\Delta^0_2=0$; then, if 
$\Delta^{k-1}_2 < \Delta^{k-1}_1$, we have
$$
\Delta^k_2 = \norm{\Phi^k_2(\Phi^{T_s}(z_\epsilon))-\Phi^k_2(\pi_\epsilon)}=
\norm{\Phi_2(\Phi^{k-1}(\Phi^{T_s}(z_\epsilon)))-\Phi_2(\Phi^{k-1}(\pi_\epsilon))}
$$
$$
\leq {1\over \lambda_u} \Delta^{k-1}_2 +
\norm{f_2(\Phi^{k-1}(\Phi^{T_s}(z_\epsilon)))-f_2(\Phi^{k-1}(\pi_\epsilon))}
$$
$$
\leq {1\over \lambda_u} \Delta^{k-1}_2 + A\eta \delta
(\Delta^{k-1}_1 + \Delta^{k-1}_2)
\leq  \Big (  {1\over \lambda_u} + 2A \eta\delta\Big )\Delta^{k-1}_1
< \Delta^{k-1}_1
$$
as soon as
$$
 {1\over \lambda_u} + 2A \eta\delta  < 1  .
$$
Therefore, we have
$$
\Delta^k_1 =  \norm{\Phi^k_1(\Phi^{T_s}(z_\epsilon))-\Phi^k_1(\pi_\epsilon)}\leq 
\lambda_u \Delta^{k-1}_1+A\eta\delta (\Delta^{k-1}_1 + \Delta^{k-1}_2)
$$
$$
\leq 
(\lambda_u+ 2\eta A \delta )\Delta^{k-1}_1 \leq 
(\lambda_u+ 2\eta A \delta )^k \Delta^{0}_1 = 
\lambda_u^k \Big ( 1+{2\eta A \delta\over \lambda_u}\Big )^k \Delta_\epsilon
\leq e \lambda_u^k   \Delta_\epsilon
$$
and
$$
\Delta^k_1 \geq \lambda_u \Delta^{k-1}_1-A\eta (\Delta^{k-1}_1 + \Delta^{k-1}_2)
\geq (\lambda_u - 2A\eta \delta )\Delta^{k-1}_1
$$
$$
\geq \lambda_u^k \Big ( 1-{2\eta A \delta\over \lambda_u}\Big )^k \Delta_\epsilon
\geq {1\over e}\lambda_u^k   \Delta_\epsilon  
$$
as soon as $k\leq T$ and
$$
 {2\eta A \delta\over \lambda_u} \leq {1\over 2 k}  .
$$
We obtained
$$
{1\over e}\lambda_u^k   \Delta_\epsilon   \leq \Delta^k_1
\leq  e \lambda_u^k   \Delta_\epsilon  .
$$
We now provide an estimate of $\Phi^k_1(\pi_\epsilon)$ and $\Phi^k_2(\pi_\epsilon)$.
We consider the segment which joins the origin $(0,0)$ and 
$\Phi^k(\pi_\epsilon)$ and define
$$
\delta_1^k=\norm{\Phi^k_1(\pi_\epsilon)}\ \ ,\ \ \delta_2^k=
\norm{\Phi^k_2(\pi_\epsilon)}  .
$$
We have $\delta_1^k < \delta_2^k$ for any $k$. In fact, since  
$\pi_\epsilon\in B(\delta)$ and $\pi_\epsilon \in W^l_s$, then 
$\Phi^k(\pi_\epsilon)\in B(\delta)$ for any $k$ and we have
$$
\delta_1^k = \norm{\Phi^k_1(\pi_\epsilon)} = 
\norm{w_s(\Phi^k_2(\pi_\epsilon))} \leq \eta 
\norm{\Phi^k_2(\pi_\epsilon)}^2\leq \eta \delta \delta^k_2< \delta^k_2 
$$
as soon as 
$$
\eta \delta < 1  .
$$
For $k=0$ we have
$$
\delta_1^0=\norm{(\pi_\epsilon)_1}=\norm{w_s((\pi_\epsilon)_2 )}\leq  
\eta \norm{\Phi^{T_s}_2(z_s)}^2 \leq \eta \delta \norm{\Phi^{T_s}_2(z_s)}\leq
\eta \delta^2 \ \ ,\ \  \delta_2^0=\norm{\Phi^{T_s}_2(z_s)} \leq \delta
$$
Then, we have
$$
\delta_2^k=\norm{\Phi^k_2(\pi_\epsilon)}=
\norm{\Phi_2(\Phi^{k-1}(\pi_\epsilon))} \leq 
{1\over \lambda_u}\norm{\Phi^{k-1}_2(\pi_\epsilon)}+ 
\eta \Norm{\Phi^{k-1}(\pi_\epsilon)}^2\leq 
{1\over \lambda_u} \delta_2^{k-1}+\eta (\delta_2^{k-1})^2
$$
$$
\leq \Big ({1\over \lambda_u} +\eta  \delta\Big )\delta_2^{k-1} 
\leq \Big ({1\over \lambda_u} +\eta  \delta\Big )^k \delta_2^{0}
\leq {1\over \lambda_u^k }\Big (1 +\eta \lambda_u  \delta\Big )^k 
\delta \leq {1\over \lambda_u^k }e\delta
$$
as soon as
$$
\eta \lambda_u  \delta \leq {1\over e k}  ,
$$
and
$$
\delta_1^k=\norm{w_s(\Phi^k_2(\pi_\epsilon)}\leq 
\eta \norm{\Phi^k_2(\pi_\epsilon)}^2= \eta (\delta_2^k)^2 
\leq \eta e{1\over \lambda_u^k }\delta  .
$$
Therefore, from
$$
{1\over e}\lambda_u^k   \Delta_\epsilon   \leq \norm{\Phi^k_1(\Phi^{T_s}(z_\epsilon))-\Phi^k_1(\pi_\epsilon)}\leq  e \lambda_u^k   \Delta_\epsilon  
$$
we have
$$
{1\over e}\lambda_u^k\Delta_\epsilon  -  \eta e{1\over \lambda_u^k }\delta\leq  
\norm{\Phi^k_1(\Phi^{T_s}(z_\epsilon))}\leq 
 e \lambda_u^k   \Delta_\epsilon +  \eta e{1\over \lambda_u^k }\delta  .
$$
Finally, from $\Delta^k_2< \Delta^k_1$ we have
$$
\norm{\Phi^k_2(\Phi^{T_s}(z_\epsilon))}\leq  \Delta^k_1+\norm{\Phi^k_2(\pi_\epsilon)}
\leq e\lambda_u^k \Delta_\epsilon +{1\over \lambda_u^k }e\delta  .
$$
From the definition of $T_\epsilon$, we have
$$
{e\delta\over \Delta_\epsilon} \leq \lambda_u^{T_\epsilon} < {\lambda_u e\delta
\over \Delta_\epsilon }  ,
$$
and therefore we have
$$
\delta - \eta \Delta_\epsilon \leq 
\delta   -  \eta e{\delta\over \lambda_u^{T_\epsilon} }\leq  
\norm{\Phi^k_1(\Phi^{T_s}(z_\epsilon))}\leq 
 e^2 \lambda_u \delta  +  \eta \Delta_\epsilon  
$$
$$
\norm{\Phi^k_2(\Phi^{T_s}(z_\epsilon))}\leq  e^2 \lambda_u \delta  +  
2\eta \Delta_\epsilon  .
$$
Therefore, since $ \Delta_\epsilon \leq (1+\lambda_w)
\lambda_\Phi^{T_s}\epsilon$, as soon as
$$
\max (1,\eta) (1+\lambda_w)\lambda^{T_s}\epsilon < \delta^2
$$
we have 
$$
\Norm{\Phi^k(\Phi^{T_s}(z_\epsilon))} \leq e^2 \lambda_u \delta  +  
2\eta \Delta_\epsilon \leq e^2 \lambda_u \delta +2 \eta (1+\lambda_w)
\lambda_\Phi^{T_s}\epsilon <  e^2 \lambda_u \delta +\delta^2 
< 2e^2\lambda_u \delta = A\delta
$$
for $A= 2e^2\lambda_u$. The thresholds conditions on $\delta$ become
$$
\delta \leq {1\over 4e^2\lambda_u\eta}\Big (1-{1\over \lambda_u}\Big )
\ \ ,\ \ 
\delta \leq {1\over 8e^2\eta}{1\over T_\epsilon}
\ \ ,\ \ 
\delta \leq {1\over 2e^3\lambda_u^2\eta}{1\over T_\epsilon}  .
$$
We now consider the point
$$
\zeta_\epsilon = \Big (\Phi^{T_s+T_\epsilon}_1(z_\epsilon),w_u(\Phi^{T_s+T_\epsilon}_1
(z_\epsilon))\Big )
$$
and the segments which join $\Phi^{-k}(\zeta_\epsilon)$ and 
$\Phi^{-k}(\Phi^{T_s+T_\epsilon}(z_\epsilon))$, for $k\leq T_\epsilon$. We already
know that $\Phi^{-k}(\Phi^{T_s+T_\epsilon}(z_\epsilon))\in B(A\delta )$, 
and 
$$
\Norm{\zeta_\epsilon }=\Norm{\Big (\Phi^{T_s+T_\epsilon}(z_\epsilon),w_u(\Phi^{T_s+T_\epsilon}(z_\epsilon))\Big )}\leq \max ( A\delta, \eta A \delta^2)\leq 
A\delta
$$
as soon as $\eta \delta \leq 1$. By definition of local unstable manifold, 
we have $\Phi^{-k}(\zeta_\epsilon) \in B(A\delta)$. We define
$$
\Delta^{-k}_1 = \norm{\Phi^{-k+T_s+T_\epsilon}_1(z_\epsilon)-
\Phi^{-k}_1(\zeta_\epsilon)}
$$
$$
\Delta^{-k}_2 = \norm{\Phi^{-k+T_s+T_\epsilon}_2(z_\epsilon) )-\Phi^{-k}_2(\zeta_\epsilon)}  ,
$$
in particular we have
$$
\Delta^{0}_1=0 \ \ ,\ \ \Delta^{0}_2 := \Delta^\epsilon  .
$$
By repeating the above arguments using the inverse map $\Phi^{-1}(x)$, 
we have $\Delta^{-k}_1 < \Delta^{-k}_2$ for any $k$ and:
$$
  {1\over e}\lambda_u^k \Delta^\epsilon \leq \Delta_2^{-k} \leq 
e\lambda_u^k \Delta^\epsilon 
$$
$$
 \Delta^\epsilon =  \norm{\Phi^{T_s+T_\epsilon}_2(z_\epsilon) -(\zeta_\epsilon)_2} 
\leq {e\over \lambda_u^{T_\epsilon}} \Delta_2^{-T_\epsilon} \leq 
 {e\over \lambda_u^{T_\epsilon}}2A\delta   .
$$
It remains to prove (\ref{vtanA2}). For any $k\leq T_\epsilon$, we have:
$$
\Norm{\Phi^k(\Phi^{T_s}(z_\epsilon))} \leq 
e \lambda_u^k \Delta_\epsilon +\max(1,\eta) e {1\over  \lambda_u^k}\delta 
\leq e^2 {\lambda_u\over \lambda_u^{T_\epsilon-k}}  \delta+
\max(1,\eta) e {1\over  \lambda_u^k}\delta
$$
and
$$
\sum_{k=0}^{T_\epsilon-1} \Norm{\Phi^k(\Phi^{T_s}(z_\epsilon))} \leq 
\sum_{k=0}^{T_\epsilon-1}  \left ( e^2 {\lambda_u\over \lambda_u^{T_\epsilon-k}}+
\max(1,\eta) e {1\over  \lambda_u^k}\right )\delta 
$$
$$
\leq  2 e \lambda_u \max(e,\eta)\delta \sum_{k=0}^{T_\epsilon} {1\over  \lambda_u^k}
\leq 2 e \lambda_u \max(e,\eta){\lambda_u \over  \lambda_u-1}\delta  .
$$
so that, by using also lemma \ref{lemma1}, we have:
$$
\Norm{D\Phi^{T_\epsilon}_{\Phi^{T_s}(z_\epsilon)} D\Phi^{T_s}_{z_\epsilon}v - 
A^{T_\epsilon} D\Phi^{T_s}_{z_\epsilon}v} \leq 
\Norm{D\Phi^{T_\epsilon}_{\Phi^{T_s}(z_\epsilon)}-A^{T_\epsilon}} \Norm{D\Phi^{T_s}_{z_\epsilon}v}
$$
$$
\leq \eta \lambda_u^{T_\epsilon} \left ( 1+ 
2\eta e^2 \delta\right )^{T_\epsilon-1} \sum_{k=0}^{T_\epsilon-1} \Norm{\Phi^k(\Phi^{T_s}(z_\epsilon))} \Norm{D\Phi^{T_s}_{z_\epsilon}v}
$$
$$
\leq \eta  \max(e,\eta){2e^2\lambda_u^2 \over  \lambda_u-1}\ \lambda_u^{T_\epsilon} \delta \Norm{D\Phi^{T_s}_{z_\epsilon}v}  .
$$
We have
$$
\Norm{D\Phi^{T_s+T_\epsilon}_{z_\epsilon}v -A^{T_\epsilon}w}\leq 
\Norm{D\Phi^{T_\epsilon}_{\Phi^{T_s}(z_\epsilon)} D\Phi^{T_s}_{z_\epsilon}v-A^{T_\epsilon}
D\Phi^{T_s}_{z_\epsilon}v}+
\Norm{A^{T_\epsilon} (D\Phi^{T_s}_{z_\epsilon}v-w)}
$$
$$
\leq 
\eta  \max(e,\eta){2e^2\lambda_u^2 \over  \lambda_u-1}\ \lambda_u^{T_\epsilon} \delta \Norm{D\Phi^{T_s}_{z_\epsilon}v}+
\Norm{A}^{T_\epsilon}\Norm{D\Phi^{T_s}_{z_\epsilon}v-w}
$$
$$
\leq 
\eta  \max(e,\eta){2e^2\lambda_u^2 \over  \lambda_u-1}\ \lambda_u^{T_\epsilon} \delta \Big ( \Norm{w}+\Norm{D\Phi^{T_s}_{z_\epsilon}v-w}\Big )+
{\lambda_u}^{T_\epsilon}\Norm{D\Phi^{T_s}_{z_\epsilon}v-w}
$$
and using (\ref{vtanA1}) and $\Norm{w}=\Norm{w_u}$ we obtain
$$
\Norm{D\Phi^{T_s+T_\epsilon}_{z_\epsilon}v  -A^{T_\epsilon}w}\leq\lambda_u^{T_\epsilon}
\left (\eta  \max(e,\eta){2e^2\lambda_u^2 \over  \lambda_u-1}\  \delta( 1+ \lambda^{T_s}\epsilon )+\lambda^{T_s}\epsilon\right )\Norm{w_u}
$$
and, since $\lambda^{T_s}\epsilon \leq \delta^2\leq \delta \leq 1$:
\begin{equation}
\Norm{D\Phi^{T_s+T_\epsilon}_{z_\epsilon}v  -A^{T_\epsilon}(w_u+w_s)}\leq\lambda_u^{T_\epsilon}
\delta \left (\eta  \max(e,\eta){4e^2\lambda_u^2 \over  \lambda_u-1}+1
\right )\Norm{w_u}  .
\end{equation}

\section{Two Technical Lemmas}\label{section5}

In this Section we prove two technical Lemmas which we obtain by 
using Lipschitz inequalities for $\Phi$ and $D\Phi$. 
\begin{lemma}\label{lemma1}
Let $U\subseteq {\Bbb R}^n$ be a neighbourhood 
of $0$, and $\Phi: U\rightarrow {\Bbb R}^n$ be a smooth map:
$$
\Phi(z)=Az+f(z)
$$
with $f_i(0,\ldots ,0)=0$, ${\partial f_{i}\over\partial z_j}(0,\ldots ,0)=
0$ for any $i,j$ and, for any $z\in B(R)$, satisfying
$$
\Norm{f(z)}\leq \eta\Norm{z}^2\ \ ,\ \ 
\Norm{Df_z} \leq \eta \Norm{z}\ \ ,\ \ \Norm{D\Phi_z}\leq l .
$$
Then, for any $z,K$ such that $\Phi^k(z)\in B(R)$ for any 
$k=0,\ldots ,K$, we have
\begin{equation}
\Norm{D\Phi^k_z - A^k} \leq  \eta \sum_{j=0}^{k-1} \lambda_u^j
\Norm{\Phi^{j}(z)}\left (\lambda_u+\eta \Norm{\Phi^{j+1}(z)}\right )\ldots 
\left (\lambda_u+\eta \Norm{\Phi^{k-1}(z)}\right )
\label{eqlemma}
\end{equation}
and
\begin{equation}
\Norm{D\Phi^k_z - A^k} \leq  \eta 
\Big (\lambda_u+ \eta \max_{j\leq k-1}\Norm{\Phi^{j}(z)}\Big )^{k-1}
\sum_{j=0}^{k-1}\Norm{\Phi^{j}(z)}  .
\label{eqlemma1}
\end{equation}
\end{lemma}
\vskip 0.4 cm
\noindent
{\bf Proof of Lemma \ref{lemma1}.} For $k=1$ we have 
$\Norm{D\Phi_z - A} = \Norm{Df_z} \leq \eta \Norm{z}$.
For generic $k\leq K$, since $\Norm{A}=\lambda_u$, we have
$$
\Norm{D\Phi^k_z - A^k} \leq \Norm{D\Phi_{\Phi^{k-1}(z)}D\Phi^{k-1}_z-A^k}
= \Norm{D\Phi_{\Phi^{k-1}(z)} (D\Phi^{k-1}_z-A^{k-1})+
(D\Phi_{\Phi^{k-1}(z)}-A)A^{k-1}}
$$
$$
\leq  \Norm{D\Phi_{\Phi^{k-1}(z)}}\Norm{D\Phi^{k-1}_z-A^{k-1}}
+\Norm{D\Phi_{\Phi^{k-1}(z)}-A}\Norm{A}^{k-1}
$$
$$
=\Norm{A+Df_{\Phi^{k-1}(z)}}\Norm{D\Phi^{k-1}_z-A^{k-1}}
+\Norm{D\Phi_{\Phi^{k-1}(z)}-A}\Norm{A}^{k-1}
$$
$$
\leq \left   (\lambda_u+\eta \Norm{\Phi^{k-1}(z)}\right )
\Norm{D\Phi^{k-1}_z-A^{k-1}}+\eta \Norm{\Phi^{k-1}(z)} \lambda_u^{k-1} .
$$ 
Assuming that (\ref{eqlemma}) is valid for $k-1$, we have
$$
\Norm{D\Phi^k_z - A^k} \leq  
\left   (\lambda_u+\eta \Norm{\Phi^{k-1}(z)}\right )
\eta \sum_{j=0}^{k-2} \lambda_u^j
\Norm{\Phi^{j}(z)}\left (\lambda_u+\eta \Norm{\Phi^{j+1}(z)}\right )\ldots 
\left (\lambda_u+\eta \Norm{\Phi^{k-2}(z)}\right )
$$
$$
+\eta \Norm{\Phi^{k-1}(z)} \lambda_u^{k-1} 
= \eta \sum_{j=0}^{k-2} \lambda_u^j
\Norm{\Phi^{j}(z)}\left (\lambda_u+\eta \Norm{\Phi^{j+1}(z)}\right )\ldots 
\left (\lambda_u+\eta \Norm{\Phi^{k-1}(z)}\right )
+\eta \Norm{\Phi^{k-1}(z)} \lambda_u^{k-1} 
$$
$$
= \eta \sum_{j=0}^{k-1} \lambda_u^j
\Norm{\Phi^{j}(z)}\left (\lambda_u+\eta \Norm{\Phi^{j+1}(z)}\right )\ldots 
\left (\lambda_u+\eta \Norm{\Phi^{k-1}(z)}\right )  .
$$
From (\ref{eqlemma}) we immediately obtain (\ref{eqlemma1}).
\vskip 0.4 cm
\noindent
\begin{lemma}\label{lemma2} 
Let $U\subseteq {\Bbb R}^n$ be a neighbourhood of $0$, and 
$\Phi: U\rightarrow {\Bbb R}^n$ be a smooth map with 
finite Lipschitz constants $\lambda_\Phi$, $\lambda_{D\Phi}$ for 
$\Phi$ and $D\Phi$ respectively. For any initial conditions $z'_0,z''_0$, 
their time--evolutions $z'_k=\Phi^k(z'_0)$, $z''_k=\Phi^k(z''_0)$ satisfy 
\begin{equation}
\Norm{z'_T-z''_T}\leq \lambda_{\Phi}^T \Norm{z'_0-z''_0}
\label{dmap}
\end{equation}
and for any $v\ne 0$, the time--evolution of the tangent vectors
$$
v'_T = D\Phi^T_{z'_0}v\ \ ,\ \ v''_T = D\Phi^T_{z''_0}v
$$
satisfies
\begin{equation}
{\Norm{v'_T-v''_T} \over \Norm{v''_T}} \leq \lambda^T \Norm{z'_0-z''_0}  .
\label{dtanmap}
\end{equation}
with
$$
\lambda = \max \left (\lambda_\Phi,  
{\Norm{D\Phi}+\lambda_{D\Phi} \over \sigma}\right ) 
$$
where $\sigma= \min_{z\in U} \min_{\Norm{v}=1}\Norm{D\Phi_zv}$. 
\end{lemma}
\vskip 0.4 cm
\noindent
{\bf Proof of Lemma \ref{lemma2}.} We prove (\ref{dmap}) by induction on $T$. 
If 
$T=1$ we have
$$
\Norm{z'_1-z''_1}=\Norm{\Phi(z'_0)-\Phi(z''_0)}\leq \lambda_\Phi \Norm{
z'_0-z''_0}  . 
$$
Le us assume
$$
\Norm{z'_{T-1}-z''_{T-1}}\leq \lambda_{\Phi}^{T-1} \Norm{z'_0-z''_0}  .
$$
Then, we have
$$
\Norm{z'_T-z''_T} = \Norm{\Phi(z'_{T-1})-\Phi(z''_{T-1})}
\leq \lambda_{\Phi} \Norm{z'_{T-1}-z''_{T-1}} \leq 
 \lambda_{\Phi}^T \Norm{z'_0-z''_0}  .
$$
Then, let us prove (\ref{dtanmap}) by induction on $T$. If $T=1$, 
we have
$$
{\Norm{v'_1-v''_1}\over \Norm{v''_1}} = 
{\Norm{(D\Phi_{z'_0}-D\Phi_{z''_0})v}\over \Norm{v''_1}}=
{\Norm{(D\Phi_{z'_0}-D\Phi_{z''_0})v}\over \Norm{v}} 
{\Norm{v}\over \Norm{v''_1}}=
{\Norm{(D\Phi_{z'_0}-D\Phi_{z''_0})v}\over \Norm{v}}
{\Norm{v}\over \Norm{D\Phi_{z''_0}v}}  .
$$
By Lipschitz estimate and inequality (\ref{sigma}) we have:
$$
{\Norm{(D\Phi_{z'_0}-D\Phi_{z''_0})v}\over \Norm{v}} \leq 
\Norm{D\Phi_{z'_0}-D\Phi_{z''_0}} \leq \lambda_{D\Phi}
\Norm{z'_0-z''_0}  
$$
$$
{\Norm{D\Phi_{z''_0}v} \over \Norm{v}} \geq 
\min_{\Norm{v}=1} \Norm{D\Phi_{z''_0}v} =\sigma >0  ,
$$
and therefore we obtain
$$
{\Norm{v'_1-v''_1}\over \Norm{v''_1}}\leq {\lambda_{D\Phi}\over \sigma}
\Norm{z'_0-z''_0}\leq \lambda \Norm{z'_0-z''_0}  .
$$
We now assume that  (\ref{dtanmap}) is satisfied for $T-1$, that is:
\begin{equation}
{\Norm{v'_{T-1}-v''_{T-1}} \over \Norm{v''_{T-1}}} \leq 
 \lambda^{T-1} \Norm{z'_0-z''_0}  .
\label{tmenuno}
\end{equation}
Then, let us consider
$$
{\Norm{v'_T-v''_T} \over \Norm{v''_T}} = 
{\Norm{D\Phi_{z'_{T-1}}v'_{T-1}-D\Phi_{z''_{T-1}}v''_{T-1}} \over 
\Norm{v''_T}} 
$$
$$
\leq {\Norm{D\Phi_{z'_{T-1}}(v'_{T-1}-v''_{T-1})}
\over \Norm{v''_T}}+
{\Norm{(D\Phi_{z'_{T-1}}-D\Phi_{z''_{T-1}})v''_{T-1}} \over 
\Norm{v''_T}}
$$
$$
\leq \Big (\sup_z\Norm{D\Phi_{z}}\Big ) 
{\Norm{v'_{T-1}-v''_{T-1}} \over \Norm{v''_T}}+
\Norm{D\Phi_{z'_{T-1}}-D\Phi_{z''_{T-1}}}{\Norm{v''_{T-1}}\over 
\Norm{v''_T}}
$$
$$
= \Big (\sup_z\Norm{D\Phi_{z}}\Big ) {\Norm{v'_{T-1}-v''_{T-1}} \over \Norm{v''_{T-1}}}
{\Norm{v''_{T-1}}\over 
\Norm{v''_T}}+\lambda_{D\Phi}\Norm{z'_{T-1}-z''_{T-1}}{\Norm{v''_{T-1}}\over 
\Norm{v''_T}}  .
$$
Using (\ref{sigma}):
$$
{\Norm{v''_{T-1}}\over 
\Norm{v''_T}} = {\Norm{v''_{T-1}}\over 
\Norm{D\Phi_{z''_{T-1}}v''_{T-1}}}\leq {1\over \sigma}
$$
and (\ref{dmap}), (\ref{tmenuno}), we obtain
$$
{\Norm{v'_T-v''_T} \over \Norm{v''_T}} \leq 
{ \Big (\sup_z\Norm{D\Phi_{z}}\Big )\over \sigma}  \lambda^{T-1} \Norm{z'_0-z''_0}
+{\lambda_{D\Phi}\over \sigma} \lambda_{\Phi}^{T-1}\Norm{z'_0-z''_0}
$$
$$
= {  \Big (\sup_z\Norm{D\Phi_{z}}\Big )\lambda^{T-1} +\lambda_{D\Phi}\lambda_{\Phi}^{T-1}\over \sigma}
\Norm{z'_0-z''_0}\leq \lambda^T\Norm{z'_0-z''_0}  .
$$
\hfill $\Box$

\section{Conclusions}

In this paper we have explained why the FLI indicators, suitably 
modified by the introduction of test functions, may be used for high 
precision computations of the stable and unstable manifolds of dynamical
systems, including the critical computations of the so called tube manifolds
of the restricted three--body problem. An advantage of the FLI method 
is that it does not requires a preliminary high precision localization of the 
hyperbolic fixed points or periodic orbits to provide high precision 
computations of their stable and unstable manifolds. This is particularly 
useful for practical applications, since additional perturbations 
can be easily included in the numerical computations.

\section*{Acknowledgments}
 Part of the computations have been done on the ``Mesocentre SIGAMM"
machine, hosted by the Observatoire de la Cote d'Azur.


\begin{thebibliography}{99}


\bibitem{Arnold64}
V.I. Arnold,
\newblock Instability of dynamical systems with several degrees of freedom. 
{ Sov. Math. Dokl.}, 6: 581--585, (1964).


\bibitem{BenGalStre76}
Benettin G. Galgani L. and Strelcyn J.M.
\newblock Kolmogorov entropy and numerical experiments. 
\newblock Physical Review A, Vol. 14, n. 6, 2338--2345, 1976.

\bibitem{clsf}
Celletti A.,  Lega E., Stefanelli L. and Froeschl\'e C.
\newblock Some results on the global dynamics of the regularized 
restricted three--body problem with dissipation. 
\newblock { Cel. Mech. and Dyn. Astr., 109, 265-284, 2011.}

\bibitem{cs}
P. Cincotta, C. Sim\'o, 
Simple tools to study global dynamics in non-axisymmetric galactic potentials - I. 
Astron. Astrophys. Sup. 147, 205 (2000).



\bibitem{FGL00}
C. Froeschl\'e, M. Guzzo  and E. Lega, 
Graphical Evolution of the Arnold Web: From Order to Chaos. 
{ Science},  289, n. 5487: 2108-2110 (2000) .


\bibitem{FGL05}
C. Froeschl\'e, M. Guzzo and E. Lega,  
\newblock Local and global diffusion along resonant lines in discrete
quasi--integrable dynamical systems. 
Cel. Mech. and Dyn. Astron., 92, 1-3: 243-255, 2005. 

\bibitem{FroLeGo96}
C.~Froeschl\'e, E.~Lega, and R.~Gonczi.
\newblock Fast Lyapunov indicators. Application to asteroidal motion.
\newblock { Celest. Mech. and Dynam. Astron.},  67: 41--62, (1997).



\bibitem{guzzoplanets}
M. Guzzo,
The web of three--planets resonances in the outer Solar System. 
Icarus, vol. 174, n. 1.,  273-284, 2005.

\bibitem{guzzoplanets2}
M. Guzzo M.,
The web of three-planet resonances in the outer solar system II: a source of orbital instability for Uranus and Neptune.
Icarus,  181, 475-485, 2006. 


\bibitem{Guzzob07}
Guzzo M.,
\newblock Chaos and diffusion in dynamical systems through stable--unstable
 manifolds, in "Space Manifolds Dynamics: Novel Spaceways for Science and
Exploration", proceedings of the conference: "Novel spaceways for scientific
 and exploration missions, a dynamical systems approach to affordable and
 sustainable space applications" held in Fucino Space Centre (Avezzano) 15--17
 October 2007. Editors: Perozzi and Ferraz Mello.  Springer. 2010. 


\bibitem{GLF02}
M. Guzzo,  E. Lega and C. Froeschl\'e,   
\newblock On the numerical detection of the effective stability of chaotic 
motions in quasi-integrable systems.  
{ Physica D}, 163, 1-2: 1-25 (2002).

\bibitem{GLF05}
M. Guzzo, E. Lega and C. Froeschl\'e,  
\newblock First Numerical Evidence of Arnold diffusion in quasi--integrable
systems. { DCDS B}, 5, 3: 687-698 (2005).

\bibitem{GLF09Fiori}
Guzzo M., Lega E. and Froeschl\'e C.,
\newblock A numerical study of the topology of normally hyperbolic invariant 
manifolds supporting Arnold diffusion in quasi-integrable systems.
Physica D, 182 , 1797--1807, 2009.


\bibitem{GLF11}
M. Guzzo M., E. Lega and C. Froeschl\'e,   
\newblock
First numerical investigation of a conjecture by N.N. Nekhoroshev about
stability in quasi-integrable systems. 
Chaos, 21, Issue 3 (2011).

\bibitem{GL13mnras}
M. Guzzo M., E. Lega,   
\newblock
On the identification of multiple close-encounters in the planar circular restricted three body problem. Monthly Notices of the Royal Astronomical Society, 
428, 2688-2694, 2013. 

\bibitem{GL13}
M. Guzzo M., E. Lega,   
\newblock
 The numerical detection of the Arnold web and its use for long-term diffusion studies in conservative and weakly dissipative systems, Chaos, vol. 23, 
023124, 2013. 


\bibitem{henheil}
H\'enon M. and Heiles C.:
\newblock The Applicability of the Third Integral of Motion: Some 
Numerical Experiments.
\newblock The Astronomical Journal, { 69}, p. 73--79, (1964). 

\bibitem{KLMR06}
Koon W.S., Lo M.W., Marsden J.E. and Ross S.D.
Dynamical Systems, the three body problem and space mission design.
Marsden Books. ISBN 978-0-615-24095-4, 2008.


\bibitem{Laskar90}
J.~Laskar.
\newblock The chaotic motion of the Solar System. A numerical estimate of the
size of the chaotic zones.
\newblock {\em Icarus}, { 88}:266--291, (1990).


\bibitem{Laskar92}
J.~Laskar, C.~Froeschl\'e, and A.~Celletti.
\newblock The measure of chaos by the numerical analysis of the fundamental
frequencies. Application to the standard mapping.
\newblock {\em Physica D}, { 56}:253, (1992).

\bibitem{Laskar93}
J.~Laskar.
\newblock Frequency analysis for multi-dimensional systems. Global dynamics and
diffusion.
\newblock {\em Physica D}, { 67}:257--281, (1993).

\bibitem{LGF03}
E. Lega, M. Guzzo and C. Froeschl\'e, 
\newblock Detection of Arnold diffusion in
Hamiltonian systems. 
{ Physica D}, 182: 179-187 (2003). 

\bibitem{LGF10b}
E. Lega, M. Guzzo and C. Froeschl\'e,
A numerical study  of the hyperbolic manifolds in a priori unstable systems. A 
comparison with Melnikov approximations. , Cel. Mech. and Dyn. Astron., 107, 
115-127, 2010.

\bibitem{LGF11}
E. Lega, M. Guzzo and C. Froeschl\'e, 
Detection of close encounters and resonances in three-body problems
through Levi-Civita regularization,  Monthly Notices of the Royal 
Astronomical Society, 418, 107-113, 2011.  

\bibitem{mf}
T.A Mitchenko and S. Ferraz--Mello, 
Resonant structure of the outer
solar system in the neighbourhood of the planets. 
A.J.  122, 474--481, 2001.

\bibitem{robutel}
P. Robutel, 
Frequency map analysis and 
quasiperiodic decompositions, in "Hamiltonian systems and Fourier 
analysis", Editor: Benest et al., 
in Hamiltonian systems and Fourier analysis, 179–198,
Adv. Astron. Astrophys., Camb. Sci. Publ., Cambridge (2005). 


\bibitem{rl}
P. Robutel and J. Laskar, 
Frequency map and global dynamics in the Solar System I.
Icarus, 152 (2001). 

\bibitem{rg}
P. Robutel and F. Gabern,
The resonant structure of Jupiter's Trojan asteroids I. Long term stability 
and diffusion. 
{ Monthly Notices of the Royal Astronomical Society.}, 372 (2006). 

\bibitem{Simo99}
C. Sim\'o, Dynamical systems methods for space missions on a vicinity of 
collinear libration points, in Sim\'o, C., editor, Hamiltonian Systems
with Three or More Degrees of Freedom (S'Agar\'o, 1995), volume 533 of
NATO Adv. Sci. Inst. Ser. C Math. Phys. Sci., pages 223--241, Dordrecht.
Kluwer Acad. Publ., (1999).

\bibitem{Szebe67}
Szebehely~V.
\newblock { Theory of orbits}.
\newblock Academic Press, New York, 1967.

\bibitem{tb}
X.Z. Tang and A.H. Boozer, Finite time Lyapunov exponent and 
advection-diffusion equation. Phys. D, 95, 3-4, 283-305 (1996). 

\bibitem{Villac08}
Villac B.F.,
\newblock Using FLI maps for preliminary spacecraft trajectory design in multi-body environments. Cel. Mech. and Dyn. Astron., 102, 29-48, 2008.

\bibitem{wmd}
B.H. Wayne, A.V. Malykh and C.M. Danforth, 
The interplay of chaos between
the terrestrial and giant planets.
Monthly Notices of the Royal Astronomical Society
Volume 407, Issue 3, September 2010, Pages: 1859-1865.


\end{thebibliography}
\end{document}